\documentclass[12pt]{JHEP3}
\input epsf.tex
\epsfclipon

\usepackage{epsfig}
\usepackage{epstopdf}
\usepackage{epsf}
\input{epsf.sty}
\usepackage{graphicx,amsmath,amssymb}
\usepackage{epsfig,multicol}

\usepackage{bbm,bm,amsmath,amssymb}
\def\blfootnote{\xdef\@thefnmark{}\@footnotetext}

\long\def\symbolfootnote[#1]#2{\begingroup%
\def\thefootnote{\fnsymbol{footnote}}\footnote[#1]{#2}\endgroup}

\newcommand{\be}{\begin{eqnarray}}
\newcommand{\ee}{\end{eqnarray}}
\newcommand{\ben}{\begin{eqnarray*}}
\newcommand{\een}{\end{eqnarray*}}

\newcommand{\bcent}{\begin{center}}
\newcommand{\ecent}{\end{center}}
\newcommand{\benum}{\begin{enumerate}}
\newcommand{\eenum}{\end{enumerate}}
\newcommand{\bdesc}{\begin{description}}
\newcommand{\edesc}{\end{description}}
\newcommand{\bitem}{\begin{itemize}}
\newcommand{\eitem}{\end{itemize}}
\newcommand{\bquote}{\begin{quote}}
\newcommand{\equote}{\end{quote}}
\newcommand{\bhalfp}{\begin{minipage}{0.45\textwidth}}
\newcommand{\ehalfp}{\end{minipage}}
\newcommand{\bhead}{\begin{center}\bf \Large}
\newcommand{\ehead}{\end{center}\bigskip}

\def\be{\begin{equation}}
\def\ee{\end{equation}}
\def\ba{\begin{eqnarray}}
\def\ea{\end{eqnarray}}

\newcommand{\roughly}[1]{\mathrel{\raise.3ex\hbox{$#1$\kern-0.85em
\lower1ex\hbox{$\sim$}}}}

\def\2pi{\left(2\pi\right)}

\def\beq{\begin{equation}}
\def\eeq{\end{equation}}
\def\bg{\begin{eqnarray}}
\def\nd{\end{eqnarray}}
\def\bea{\begin{eqnarray}}
\def\eea{\end{eqnarray}}

\def\D3{\overline{\mbox{D3}}}

\title{Toward Large $N$ Thermal QCD from Dual Gravity: The Heavy Quarkonium Potential}

\author{Mohammed Mia, Keshav Dasgupta, Charles Gale, Sangyong Jeon\\
Ernest Rutherford Physics Building, McGill University,\\ 3600 University
Street, Montr{\'e}al QC, Canada H3A 2T8 }
\date{November 2009}

\abstract{We continue our study on the gravity duals for strongly coupled large $N$ QCD with fundamental flavors
both at zero and non-zero 
temperatures. The gravity dual at zero temperature captures the logarithmic runnings of the coupling constants 
at far IR and the almost conformal, albeit strongly coupled, behavior at the UV. The full UV completion
of gauge theory is accomplished in the gravity side by attaching an AdS cap to the IR geometry described in our 
previous work. Attaching such an AdS cap is highly non-trivial because it amounts to finding the right 
interpolating geometry and sources
that take us from a gravity solution with non-zero three-form fluxes to another one that has 
almost vanishing three-form fluxes. In this paper we give a concrete realisation of such a scenario, completing the 
program advocated in our earlier paper. One of the main advantage of having such a background, in addition to 
providing a dual description of the required gauge theory, is the absence of Landau poles and 
consequently the UV divergences of 
the Wilson loops. The potential for the heaviest fundamental quark anti-quark pairs, which are like the 
heavy quarkonium states in realistic QCD, can be computed and their linear behavior at large separations and zero temperature  could be
demonstrated. At
small separations the expected Coulombic behavior appears to dominate. 
On the other hand, at non-zero temperatures 
interesting properties like heavy quarkonium
type suppressions and melting are shown to emerge from our gravity dual. 
We provide some discussions of the melting temperature and compare our results with the 
Charmonium spectrum and lattice simulations. We argue that, in spite of the large $N$ nature of our construction, certain 
model-independent predictions can be made.}

\maketitle

\begin{document}

\section{Introduction}

The study of strongly interacting matter under extreme conditions of temperature and/or density is one of the most fascinating areas of contemporary subatomic physics. This program aims to explore the many facets of the {\it bulk} behavior of Quantum ChromoDynamics (QCD): The theory of the strong interaction. It seeks to map out the different phases allowed by QCD and the nature of the possible phase transitions connecting them or in short, to elucidate the QCD phase diagram. In this context, the existence of an exotic phase of QCD, a quark-gluon plasma, has been a prediction of lattice QCD whose details have continuously being refined over the years \cite{Bazavov:2009zn}. On the experimental front, several observables have been put forward as signature of the quark gluon plasma. These include electromagnetic radiation \cite{Gale:2009gc}, the quenching of energetic QCD jets \cite{Gyulassy:2003mc}, and the dissolution (with increasing collision centrality and energy) of heavy quark bound states according to the seminal suggestion in  Ref. \cite{Matsui:1986dk}. The Relativistic Heavy Ion Collider (RHIC), now at the end of its first decade of operation, has uncovered an intriguing set of phenomena suggestive of new physics. One of these is the observation of strong hydrodynamic flow effects, highly suggestive of a ``strongly coupled'' quark gluon plasma \cite{Gyulassy:2004zy}. 

The fate of quarkonium is being analyzed at RHIC, as it was at the SPS before. A surprising fact to emerge of these studies is that the suppression of the $c \bar{c}$ ground state - the $J/\psi$ - at RHIC is entirely comparable to that at the SPS, in spite of the much larger energy densities being reached at the first facility. This triggered many analyses with scenarios where the enhanced dissociation at RHIC was roughly compensated by  an extra formation owing, for example,  to quark-antiquark coalescence near hadronization \cite{coal}. Related investigations are concerned by the fate of the quarkonium spectral density above $T_c$, the deconfinement temperature \cite{Mocsy:2009ca}. It is fair to write that the study of quarkonium imbedded in a finite-temperature strongly interacting medium is a flourishing industry: The modifications of its spectral profile can be related to in-medium effects. A related topic of investigation on the lattice consists of calculating the quark-antiquark potential as a function of temperature \cite{lat_T}. These calculations show a Coulomb potential at zero temperature, with an added linear part that  slowly disappears as $T$ is raised, leading eventually to the unbinding of quarkonia bound states. Our goal in this work is to approach this softening of the potential from a different point of view.

In parallel with the studies described in the previous paragraph, the physics of hot and dense strongly interacting matter, and thus that of the quark-gluon plasma, has recently benefited from the use of a new set of techniques, germane to string theory. The gauge-string duality can indeed provide a sophisticated toolbox with which to treat strongly-coupled, strongly interacting systems \cite{Mal-1,Witt-1,son-1}. Our purpose here is to bring closer the more traditional investigations in QCD with those pursued in string theory. In gauge-string duality, a finite-temperature medium is dual to a black hole. Even though in a large number of applications the associated field theory is conformal, we use a framework which is ``QCD-like''. More specifically, we construct the dual gravity of thermal field theory which becomes almost conformal in the UV but has 
logarithmic running of coupling in the IR  with  matter in the fundamental representation. Without being explicitly  QCD, this string theory will provide some of the features associated with large $N$ Quantum Chromodynamics, and its study may shed more light on the behavior of strongly coupled, strongly interacting matter at finite temperature. 

Our paper is organized as follows: In the next section we define the geometry in which our solution will exist. 
The description of the full geometry is subtle, so we will divide the geometry in three regions. The far IR geometry 
will be described in sec. 2.1, and the far UV geometry will be described in sec. 2.3. These two geometries are 
connected
by an interpolating geometry that we will describe in details in sec. 2.2. Once we have the full geometry,  
we compute the heavy quark potential from the Nambu-Goto action first for zero, and then for finite temperature in 
sec. 3. In this section we will also provide a generic argument for confinement both for zero and non-zero temperatures.
Although most of our analysis in this paper will be done analytically, we will do some detailed numerical analysis to
study regimes that are difficult to access analytically. We will show that the numerical analysis fits consistently with 
the expected behavior of the heavy quarkonium states in this theory.    
Finally, we summarize and conclude.

\section{Construction of the Geometry}

Following the development in \cite{KS}, in \cite{FEP}  it was shown that a geometry where the dual thermal field theory was almost conformal in the UV, and had a logarithmic running of the coupling in the IR existed. The gauge theory studied in \cite{FEP} had a dual weakly coupled gravity description at zero temperature in terms
of a warped deformed conifold with seven branes and fluxes. The gauge theory in turn is strongly coupled with a 
{\it smooth} RG flow but no well defined colors at any given scale. When the gauge theory is weakly coupled, the 
description can be presented in terms of cascades of Seiberg dualities that slows down quite a bit when one approaches
the far IR because of the presence of fundamental flavors. There is no supergravity dual description available for this 
case, and the cascade is only captured by the full string theory on the relevant geometry. 

Once a non-zero temperature is switched on, the strongly coupled gauge theory description is given by a dual supergravity
solution on a {\it resolved} warped deformed conifold with seven branes and fluxes \cite{FEP}. The resolution factor is
directly related to the temperature because in the presence of a black hole a consistent solution of the system 
can only be achieved by introducing a non-zero resolution factor for the two cycle. 
In a Klebanov-Tseytlin type geometry, this resolution
factor would in fact remove the naked singularity. 

One of the other key ingredient of the solution presented in \cite{FEP} is the far UV picture. In all the previous 
known attempts to this problem, the dual supergravity solution was always afflicted by the presence of Landau Poles. 
Such problems arose because of the behavior of the axio-dilaton, that typically blows up due to their logarithmic 
behavior. What we pointed out in \cite{FEP} is that the logarithmic behavior, which is 
so ubiquitous in these constructions, appears because we are studying the 
theory near any one of the seven branes. In the full F-theory picture the large $r$ behavior is perfectly finite, 
and in fact also has a good description in terms of the metric too. The behavior of the 
warp factor for large $r$ is 
given by:
\bg\label{larger}
h ~= ~ \sum_{\alpha} {L^4_{(\alpha)}\over r^4_{(\alpha)}}
\nd
with $r_{(\alpha)} = r^{1+{\epsilon_{(\alpha)}\over 4}}$ and  $\epsilon_{(\alpha)}$ 
is a small positive number that is a function 
of $g_sN_f, g_sM$ and $g_sN$ (see eq. (3.36) of \cite{FEP}. The sign difference from \cite{FEP}
is just a matter of convention). 
The log $r$ appearing in the warp factor doesn't create 
much of a problem at UV: the theory is perfectly holographically renormalisable, and any fluctuations of the 
background are under control. The fact that we can have well defined and renormalisable interactions in this background
in the presence of fundamental flavors was shown, we believe for the first time, in \cite{FEP} (see \cite{AB} for 
renormalisability argument without fundamental flavors). 
For the present 
purpose, we want to ask a slightly different question here, namely: can we construct a dual supergravity background that 
allows logarithmic RG flow in the IR but has a vanishing beta function at far UV? From our discussion of the UV 
caps in \cite{FEP} it is clear what we should be looking for: we need a gravitational background that resembles 
OKS geometry for small $r$, but has a UV cap given by an asymptotic AdS geometry. To extend this configuration 
to
high temperature, we need OKS-BH geometry\footnote{For more details on the construction of OKS-BH 
(Ouyang-Klebanov-Strassler-Black-Hole) geometry, see \cite{FEP} sections 3.1 and 3.3.} 
at small $r$, and asymptotic AdS-Schwarchild geometry at large $r$. 
   
Such a geometry looks complicated, so we may want 
to ask whether we can switch off the three form fluxes and still have a dual 
description with running couplings. If this were possible then the analysis could be made much more simpler. It turns 
out however that  
such a simplification cannot occur in our set-up. To elucidate the last point, let us give a brief discussion.

The RG runnings of the two gauge groups in this theory are determined by     
the following dual maps
in terms of the bulk axio-dilaton $\tau$ and NS potential $B$:
\bg\label{maps}
&&{4\pi^2\over g_1^2} + {4\pi^2\over g_2^2} = \pi~ {\rm Im}~\tau\nonumber\\
&& {4\pi^2\over g_1^2} - {4\pi^2\over g_2^2} = {{\rm Im}~\tau\over 2\pi\alpha'} \int_{S^2} B - \pi~{\rm Im}~\tau
~~({\rm mod}~2\pi)
\nd
Once we switch off $B$ the two couplings would be the same and would induce a Shifman-Vainstein $\beta$-function of the 
form:
\bg\label{svbeta}
{\partial\over \partial ~{\rm log}~\Lambda} {8\pi^2\over g^2_{\rm YM}} ~ = ~ 3N - 2N(1-\gamma_{A, B}) -N_f(1-\gamma_q)
\nd
where $\Lambda$ is the energy scale that is related to the radial coordinate $r$ in the gravity side, and 
$\Gamma_{A,B}$ and $\gamma_q$ 
are the anomalous dimensions of bi-fundamental and fundamental fields respectively. With
such a picture of the flow, we might think that the F-theory completion might be to simply add sufficient number of seven
branes parallel to the spacetime directions and wrapping the two internal two-spheres (so that they are points in the 
($r, \psi$) plane). This simple picture would unfortunately be inconsistent with the underlying cascading 
dynamics as could be seen from a T-dual framework \cite{dasmukhi, ouyang},
and therefore would be incapable of showing certain important behavior expected from this model.

To understand the problem, observe that 
in the T-dual picture {\it \'a la} 
\cite{dasmukhi}, the D3-D7-conifold geometry is mapped to a configuration of 
D4-D6 and intersecting NS5 branes. The NS5 branes, that are 
T-dual to the conifold, are along $x^{012345}$ and $x^{012389}$ directions and have $N$ D4 branes (T-dual of $N$ D3 
branes) between them. Everytime we cross the NS5 branes we expect extra D4 branes
 to appear because of the D6 branes. This is 
however only possible if the D6 branes are along $x^{0123457}$ which in turn
would imply that in the brane side the D7 branes have to be along the 
radial direction. Additionally,
motion of the NS5 brane would imply a $B_{NS}$ field in the brane side that is not a constant but has
at least a log $r$ dependence along the radial direction. This means that $H_{NS}$ is non-zero, and 
we need to switch on $H_{RR}$ to satisfy the equations of motion, 
bringing us back to the model originally advocated in \cite{FEP}! 

The discussion above should convince the readers that there aren't much avenue to simplify the original proposal of 
\cite{FEP}. The original model proposed in \cite{FEP} is structurally complicated, but is possibly the {\it simplest} 
in realising some of the properties of IR large $N$ QCD. A model simpler than this would be deviod of any interesting 
physics.

Once this is settled, we want to see how to construct the kind of geometry that we mentioned above. Our requirement is 
to impose confinement at far IR and vanishing beta function at far UV. Since the original model studied in \cite{FEP} 
doesn't quite have the right large $r$ behavior because the warp factor therein goes as \eqref{larger}, we need
to add appropriate UV cap. However before we actually go about constructing the background, let us clarify how addition
of UV caps in general can change IR geometries. One thing should of course be clear, the far IR geometries 
{\it cannot} change by the addition of UV caps. This is because the UV caps corresponding to adding non-trivial 
irrelevant operators in the dual gauge theory\footnote{Assuming of course that the {\it relevant} operators were 
responsible for creating the cascading dynamics from a given UV completion in the first place!}. 
These operators keep far IR physics completely unchanged, but 
physics at not-so-small energies may change a bit. So the question is how are these changes registered in our 
analysis? Additionally we may also want to ask how entropy of our gauge theories affected by the addition of UV caps?

Both the above questions may be answered if we could figure out how the UV caps affect the energy momentum tensors of 
our gauge theories. The generic form of the energy-momentum tensor that we derived in our earlier paper \cite{FEP}
can be reproduced as:
\bg\label{wakeup}
&& T^{mm}_{{\rm medium} + {\rm quark}}    
 = \int \frac{d^4q}{(2\pi)^4}\sum_{\alpha, \beta}
\Bigg\{({H}_{\vert\alpha\vert}^{mn}+ {H}_{\vert\alpha\vert}^{nm})s_{nn}^{(4)[\beta]} 
-4({K}_{\vert\alpha\vert}^{mn}+ {K}_{\vert\alpha\vert}^{nm})s_{nn}^{(4)[\beta]}\nonumber\\
&& ~~~~~~~~~~~ +({K}_{\vert\alpha\vert}^{mn}+ {K}_{\vert\alpha\vert}^{nm})s_{nn}^{(5)[\beta]}
+\sum_{j=0}^{\infty}~\hat{b}^{(\alpha)}_{n(j)} \widetilde{J}^n  
\delta_{nm}  e^{-j{\cal N}_{\rm uv}} + {\cal O}({\cal T} e^{-{\cal N}_{uv}})\Bigg\}
\nd 
where ${H}_{\vert\alpha\vert}^{mn}$ and ${K}_{\vert\alpha\vert}^{mn}$ depend on the full background geometry 
via eq. (3.124) in \cite{FEP} with $s_{nn}^{(p)[\beta]}$ being the Fourier coefficients. The other terms
namely $\hat{b}^{(\alpha)}_{n(j)}$ and 
${\cal N}_{\rm uv}$ together specify the boundary theory for a specific
UV completion \cite{FEP}. 

Now its easy to see how the UV caps would change our results. Once we add a UV cap the local region 
$r_c - \alpha_1 \le r \le r_c + \alpha_2$ near the junction\footnote{Clearly $\alpha_1 << r_c$ because the far IR 
geometry should remain completely unaltered.} 
at $r = r_c$ changes, with ($\alpha_1, \alpha_2$) being some appropriate neighborhood around $r_c$. This means that 
$C_1^{mn}, A_1^{mn}$ and $B_1^{mn}$ etc. in eq (3.124) of \cite{FEP} would change. These changes can be registered as
\bg\label{changela}
&&{H}_{\vert\alpha\vert}^{mn} ~\to ~{\widetilde H}_{\vert\alpha\vert}^{mn} ~\equiv ~ {H}_{\vert\alpha\vert}^{mn} ~+~ 
(\delta C_{1(\alpha)}^{mn} - \delta A'^{mn}_{1(\alpha)})e^{-4\left[1 - \epsilon_{(\alpha)}\right]{\cal N}_{\rm eff}}
 ~+~ {\cal O}(e^{-j{\cal N}_{\rm uv}})
\nonumber\\
&&{K}_{\vert\alpha\vert}^{mn} ~ \to ~ {\widetilde K}_{\vert\alpha\vert}^{mn} ~\equiv~ {K}_{\vert\alpha\vert}^{mn} ~+~ 
(\delta B_{1(\alpha)}^{mn} - \delta A^{mn}_{1(\alpha)}) e^{-4\left[1 - \epsilon_{(\alpha)}\right]{\cal N}_{\rm eff}} 
~+~ {\cal O}(e^{-j{\cal N}_{\rm uv}})
\nd
where the last terms in both the above equations appear from additional UV degrees of freedom,
$C_{1(\alpha)}^{mn}$ etc are the relevant $\alpha$-th components of $C_{1}^{mn}$ etc, and ${\cal N}_{\rm eff}$ is the 
effective number of degrees of freedom at the cutoff. 

Once we know these changes, its not too difficult to figure out the changes in the entropies
due to the addition of UV caps. All we need are the RHS of eq (3.220) in \cite{FEP} using the results from 
\eqref{changela} and taking care of the boundary temperatures $T_b$ from the changes in the warp factors\footnote{There 
may be interesting cases where the changes in the energy-momentum tensors are compensated by the changes in the 
boundary temperatures. In such cases the entropies may remain unchanged. Here we will not consider such cases.}. 
Using \eqref{changela} the result can be written as:
\bg\label{entropych}
{\delta s\over s}  ~&= &~ \left({1\over {\cal T}} + {1\over 2h({\cal T})}{dh({\cal T})\over d{\cal T}}\right) 
\delta{\cal T} \\
&& ~ + {\int d^4 q \sum_{\alpha, \beta}\left[\delta H^{(mn)}_{\vert\alpha\vert} {\tilde s}_{nn}^{(4)[\beta]} 
-\delta K^{(mn)}_{\vert\alpha\vert}\left(4 {\tilde s}_{nn}^{(4)[\beta]} - {\tilde s}_{nn}^{(5)[\beta]}\right)\right] 
\over 
\int d^4 q' \sum_{\alpha, \beta}\left[H^{(mn)}_{\vert\alpha\vert} {\tilde s}_{nn}^{(4)[\beta]} 
- K^{(mn)}_{\vert\alpha\vert}\left(4 {\tilde s}_{nn}^{(4)[\beta]} - {\tilde s}_{nn}^{(5)[\beta]}\right) + 
{\cal O}(e^{-{\cal N}_{uv}})\right]}\nonumber
\nd
However physics that are only sensitive to far IR dynamics of our theory will not be affected by the addition of 
UV caps. On the other hand in all cases, far IR or not, none of our results could depend on the cut-off $r_c$. The 
results are only sensitive to the changes in IR geometries (via \eqref{changela}) and the UV degrees of 
freedom (via $e^{-j{\cal N}_{\rm uv}}$). 

From the above discussions we see how IR geometries could be affected by the addition of UV caps. This then tells us that 
we cannot simply add an AdS geometry at $r = r_c$. The vanising beta function at UV could be realised by an 
asymptotic AdS geometry, a geometry whose warp factor behave as $r^{-4}$ only asymptotically. In other words we require:
\bg\label{hde}
&& h ~= ~\frac{L^4}{r^4}\left[1+\sum_{i=1}^\infty\frac{a_i(\psi,\theta_j,\phi_j)}{r^i}\right]~~~ {\rm  for ~~ large} ~r\nonumber\\
&& h ~= ~\frac{L^4}{r^4}\left[\sum_{j,k=0}\frac{b_{jk}(\psi,\theta_i,\phi_i){\rm log}^kr}{r^j}\right]~~~{\rm  for ~~ small} ~r
\nd
where ($\theta_i, \phi_i, \psi$) are the coordinates of the internal space. Observe also that we are 
now identifying the small $r$ behavior of 
the warp factor to the relation \eqref{larger} given above. The precise connection will be spelled out 
in details below.   

Let us now make this a bit more precise. We require a gauge theory with confining IR dynamics and almost free 
UV dynamics at zero temperature, and then we want to study this theory at a temperature higher 
than the deconfining temperature, as mentioned before. Our dual gravity background that could in principle reproduce the 
gauge theory dynamics couldn't be the pure OKS (or OKS-BH) background of \cite{FEP}. We need an 
appropriate UV cap. Again, as we mentioned earlier, the UV cap should be asymptotically AdS. The warp factor 
should have the form \eqref{hde} at UV and IR, so we need an interpolating geometry between them to have
a well defined background. The logarithmic warp factor at far IR tells us that the geometry is influenced by one or a set 
of coincident D7 branes.    
These seven branes wrap the  $T^{1,1}$ as in {\it branch 2} of \cite{FEP} 
while extending in the radial $r$ 
direction and filling up four Minkowski directions (see eq (3.9) of \cite{FEP}). 
In particular the embedding equation for a D7 brane is given by \cite{ouyang, FEP}
\bg \label{embedding}
z\equiv ~r^{3/2} e^{i(\psi-\phi_1-\phi_2)}{\rm sin}~\frac{\theta_1}{2}~{\rm sin}~\frac{\theta_2}{2}~=\mu
\nd
where $\mu$ is a parameter. For supersymmetric case $\mu$ would be related to the deformation parameter 
of the conifold. Since we don't require supersymmetry we can take $\mu$ to be 
arbitrary\footnote{The issue of supersymmetry is a little subtle here. The susy can of course be broken by choosing 
a different $\mu$, but can also be broken by choosing the right $\mu$ but separating the wrapped D5 branes along 
($\theta_2, \phi_2$) directions. One may say that if we allow bound states of D5 and D7 branes we might restore zero 
temperature susy. Alternatively we can consider the seven-branes to be oriented as in \cite{gtpapers} which is related to 
our far IR configuration. In \cite{gtpapers} heavy fundamental quarks could still restore susy. In this paper we will not 
consider the seven-brane configurations of \cite{gtpapers}.}. 
Different values of 
$\mu$ will tell us how far the D7 branes are from the origin $r = 0$. 
The $N_f$ D7 branes may have $N_f$
different locations given by $N_f$ different values of $\mu$ or D7 branes may be coincident with just a single value of
$\mu$. The positions of the seven branes are therefore parametrised by the coordinate $z$.
Since the seven branes are in {\it branch 2} of \cite{FEP} their positions can be precisely parametrised by the 
internal coordinates ($\theta_2, \phi_2$). Thus the seven branes stretch along $r$ and can be placed at 
any point on the ($\theta_2, \phi_2$) plane\footnote{In actual case the embedding is a union of branch 1 and branch 2.
Therefore the seven-branes will trace a complicated surface in ($r, \theta_i, \phi_i, \psi$) plane. For simplicity 
we will assume the embedding to be given by branch 2 of \cite{FEP}. Later on when we study fluxes, the non-trivial 
nature of the seven-brane embeddings will become important.}. 
Because of this distribution, as we shall see shortly, axion-dilaton
field runs with the coordinate $z$ and the running is determined by F theory.  

Our UV cap, in the full F-theory picture, should allow a distribution of seven branes that could eventually 
reproduce the warp factors \eqref{hde}. This is however
not the {\it only} requirement: we also want to study the potential 
of heavy quarkonium type bound states in our theory (this means that we need to study the bound states of very heavy 
quark-antiquark pairs). Which in turn implies that we  require a set of seven branes as far away from the origin as 
possible (or, at high temperature, as far away as possible from the black hole horizon). 
There are a few possible ways to distribute the seven branes that might be able to reproduce the required picture. 
The simplest way would be to distribute the seven branes as in Figure 1 below. 
\begin{figure}[htb]\label{sevenbraneconf1}
		\begin{center}
\includegraphics[height=6cm]{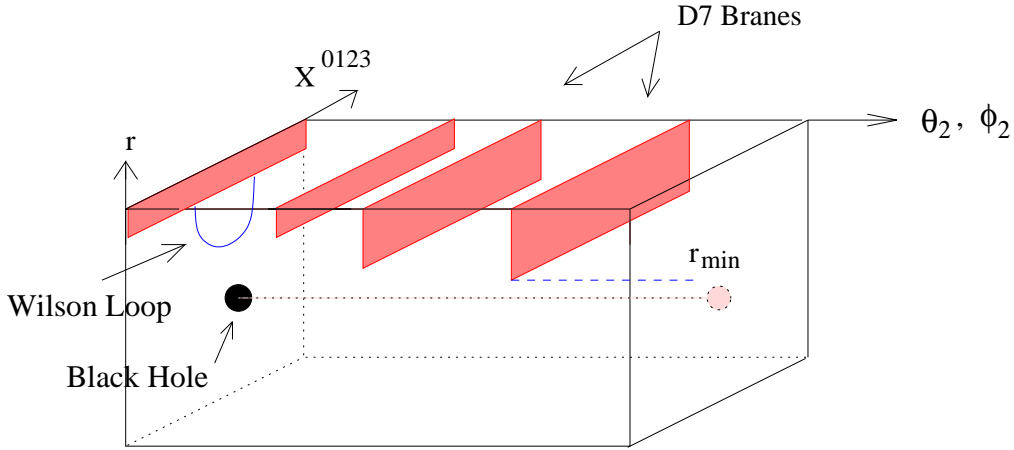}
		\caption{Simplest way to distribute localised seven branes in our model. The seven branes wrap the 
internal sphere parametrised by ($\theta_1, \phi_1$) and are stretched along the spacetime directions. Their extensions
along the radial directions are parametrised by $\mu$ as in embedding equation above. The coordinate $r_{\rm min}$ 
denote the distance of the nearest seven brane from the black hole horizon. A string stretched between this 
seven brane and the black hole horizon is the lightest {\it fundamental} quark in our model. 
The heaviest quark, on the other hand, will 
be from the seven brane that is farthest from the horizon. A string whose two ends lie on such a seven brane will 
form a quark antiquark bound state. The temporal evolution of such a string will determine the Wilson loop in 
our picture.}
		\end{center}
		\end{figure} 
This picture, although simple and desirable, however
does not quite suffice for us because we need a configuration of seven 
branes that could interpolate between the IR and UV configurations. One of the simplest way to have an interpolating 
geometry using the configurations studied in \cite{FEP} is to make the seven branes delocalised along the ($r, \theta_2, 
\phi_2$) directions and call the resulting quantity as ${\widetilde N}_f(r, \theta_2, \phi_2)$. 
This means that 
\bg\label{nfcon}
N_f(r) ~\equiv~ \int d\theta_2 d\phi_2 ~{\widetilde N}_f(r, \theta_2, \phi_2)~ 
{\rm sin}~\theta_2
\nd
An immediate way to realise such a 
configuration is given in  Figure 2 below. Such a configuration has been advocated in some 
recent works (see for example \cite{cotrone, cotrone2} where the delocalised seven branes are embedded via the Kuperstein 
embeddings \cite{kuperstein}). 
\begin{figure}[htb]\label{sevenbraneconf2}
		\begin{center}
\includegraphics[height=6cm]{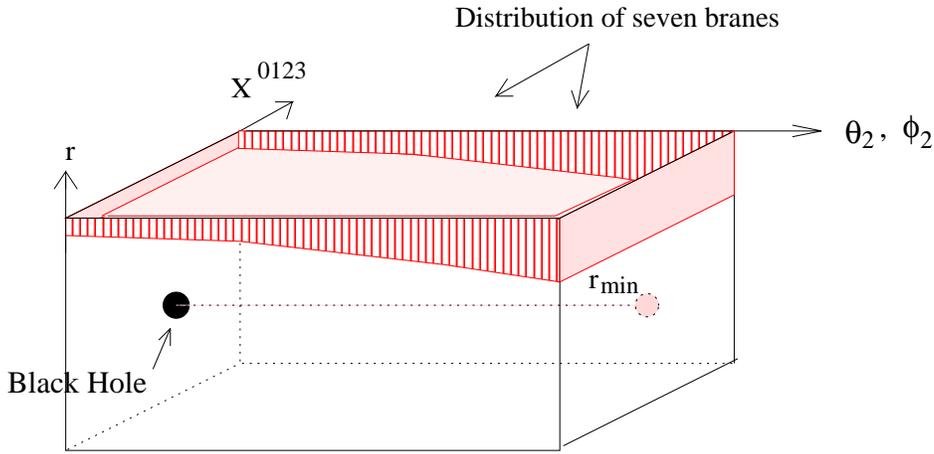}
		\caption{Complete delocalisation of the seven branes along ($r, \theta_2, \phi_2$) directions.}
		\end{center}
		\end{figure} 
Such a configuration of seven branes, although useful for many purposes, unfortunately
still does not quite suffice for us because heavy 
quarks in such a scenario would tend to go to configurations of lighter quarks spontaneously. 
Furthermore we want to impose the 
F-theory constraint, for scales $r > {\hat r}$:
\bg\label{nfconstraint}
N_f(r)\Big\vert_{r > {\hat r}}~= ~ 24
\nd
which would be a little difficult to impose in the fully delocalised scenario\footnote{In fact F-theory {\it can} allow 
number of seven-branes to be arbitrarily large. For this case we need to carefully study the singularity structure 
of the underlying manifold. Here, for most of the paper, we will restrict ourselves to $24$ seven-branes. This means that 
$g_s$ could be as small as 0.042.}. 
Therefore the configuration that we 
would be mostly interested in is given in  Figure 3. In this picture, which should be viewed as a 
cross between the earlier two figures, every individual set of seven branes are delocalised a little bit.  
\begin{figure}[htb]\label{sevenbraneconf3}
		\begin{center}
\includegraphics[height=6cm]{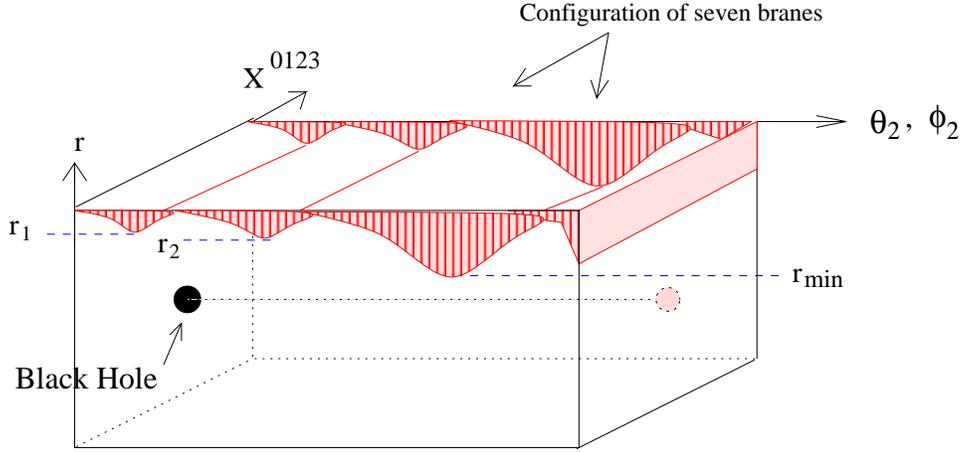}
		\caption{{Another configuration of seven branes where the delocalisation is milder compared to the 
earlier picture. The local minima of every set of seven branes help us to study various configurations of 
quark antiquarks pairs.}}
		\end{center}
		\end{figure} 
The F-theory constraint on the number of flavors i.e \eqref{nfconstraint} can be easily imposed without making 
$N_f(r, \theta_2, \phi_2)$ arbitrarily small. The final picture that we want to emphasise which would capture the 
underlying dynamics is given as  Figure 4 below. The figure is a slight variant of the previous figure. We have 
divided our geometry into three regions of interest: Regions 1, 2 and 3. 
\begin{figure}[htb]\label{sevenbraneconf4}
		\begin{center}
\includegraphics[height=6cm]{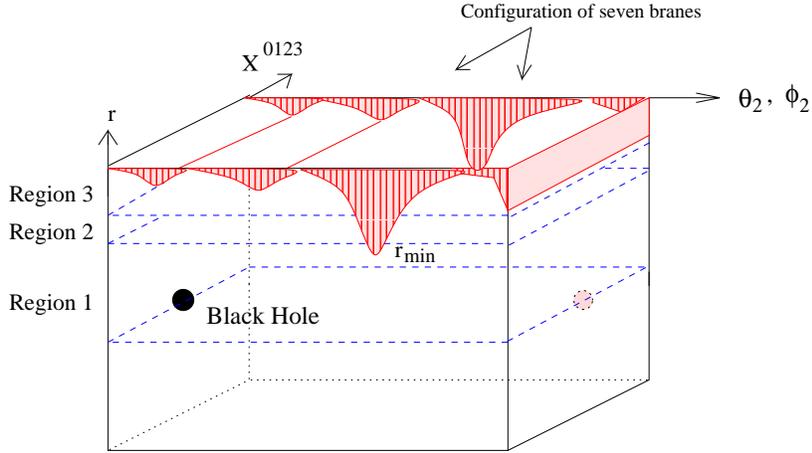}
		\caption{{This figure, which is a slight variant of the previous figure, shows the various regions
of interest. As should be clear, most of the seven branes lie in Region 3, except for a small number of 
coincident seven branes that dip till $r_{\rm min}$ i.e Region 1. The interpolating region is Region 2. The detailed 
backgrounds for each of these regions are given in the text. Note however that, although we have emphasised Region 1
more, we will only consider the case where Region 3 $>>$ Region 1 + Region 2.}}
		\end{center}
		\end{figure} 
Region 1 is basically the one discussed in great details in \cite{FEP}. In this region there is one (or a coincident 
set of) seven brane(s). The logarithmic dependences of the warp factor and fluxes come from these coincident 
(or single) seven branes. In fact the logarithmic runnings of the gauge theory coupling constants also stem from these
seven branes. 

Since we require UV free (or more appropriately, strongly coupled and conformal) the IR logarithmic runnings wouldn't 
be very desirable. Therefore the UV cap in the full F-theory framework is depicted as Region 3 in the above figure. 
In this region we expect all the seven branes to be distributed so that axio-dilaton has the right behavior. We also 
expect vanishing $H_{NS}$ and $H_{RR}$ fields (just like the AdS cases). 

It is clear that one cannot jump from Region 1 to Region 3 abruptly. There should be an interpolating geometry 
where fluxes and the metric should have the necessary property of connecting the two solutions. 
This is Region 2 in our figure above. For all 
practical purposes, we expect Region 3 to dominate, in other words, Region 3 should be greater than both 
Regions 1 and 2 combined together. In such a scenario analysis of Wilson loop for heavy quark - antiquark bounds 
states would be easy: we wouldn't have to worry too much about the intermediate regions. Another big advantage 
about our UV cap is related to the issues raised in \cite{cschu}. Since the $H_{NS}$ and the 
axio-dilaton fields have  well defined 
behaviors at large $r$, there would be {\it no} UV divergences of the Wilson loops in our picture! 
Therefore our configuration can not only boast of  
holographically renormalisability, but also of the absence of Landau poles and the associated UV divergences
of the Wilson loops.   

In the following, let us therefore discuss the backgrounds for all the three regions in details.

\subsection{Region 1: Fluxes, Metric and the Coupling Constants Flow}

The background for Region 1 is discussed in details in \cite{FEP}, so we will be brief. All the logarithmic behaviors 
for the fluxes and metric come from the single set of seven branes. The metric has the following typical form:
\bg\label{bhmet1}
ds^2 = {1\over \sqrt{h}}
\Big[-g_1(r)dt^2+dx^2+dy^2+dz^2\Big] +\sqrt{h}\Big[g_2(r)^{-1}dr^2+ d{\cal M}_5^2\Big]
\nd  
where $g_i(r)$ are the black-hole factors and $d{\cal M}_5^2$ is the metric of warped resolved-deformed conifold (see
the form in eq (3.5) of \cite{FEP}). The 
internal space retains its resolved-deformed conifold form upto ${\cal O}(g_sN_f)$. Beyond this order the 
internal space loses its simple form and becomes a complicated non-K\"ahler manifold. The warp factor to this 
order, in terms of $N^{\rm eff}_f, M_{\rm eff}$ (see eq (3.10) of \cite{FEP} for details), is:
\bg \label{hvalue}
h =\frac{L^4}{r^4}\Bigg[1+\frac{3g_sM_{\rm eff}^2}{2\pi N}{\rm log}r\left\{1+\frac{3g_sN^{\rm eff}_f}{2\pi}\left({\rm
log}r+\frac{1}{2}\right)+\frac{g_sN^{\rm eff}_f}{4\pi}{\rm log}\left({\rm sin}\frac{\theta_1}{2}
{\rm sin}\frac{\theta_2}{2}\right)\right\}\Bigg]\nonumber\\
\nd
 As discussed in \cite{FEP}, the background has {\it all} the type IIB fluxes 
switched on, namely, the three-forms, five-form and the axio-dilaton. Both $N^{\rm eff}_f$ and $M_{\rm eff}$ are 
different from $N_f$ and $M$. We will give detailed reason for this when we discuss the full geometry in the next two 
subsections. 
The three-form fluxes are:
\begin{eqnarray}\label{koremi}
{\widetilde F}_3 & = & 2M_{\rm eff} {\bf A_1} \left(1 + {3g_sN^{\rm eff}_f\over 2\pi}~{\rm log}~r\right) ~e_\psi \wedge 
\frac{1}{2}\left({\rm sin}~\theta_1~ d\theta_1 \wedge d\phi_1-{\bf B_1}~{\rm sin}~\theta_2~ d\theta_2 \wedge
d\phi_2\right)\nonumber\\
&& -{3g_s M_{\rm eff}N^{\rm eff}_f\over 4\pi} {\bf A_2}~{dr\over r}\wedge e_\psi \wedge \left({\rm cot}~{\theta_2 \over 2}
~{\rm sin}~\theta_2 ~d\phi_2 
- {\bf B_2}~ {\rm cot}~{\theta_1 \over 2}~{\rm sin}~\theta_1 ~d\phi_1\right)\nonumber \\
&& -{3g_s M_{\rm eff}N^{\rm eff}_f\over 8\pi}{\bf A_3} ~{\rm sin}~\theta_1 ~{\rm sin}~\theta_2 \left({\rm cot}~{\theta_2 \over 2}
~d\theta_1 +
{\bf B_3}~ {\rm cot}~{\theta_1 \over 2}~d\theta_2\right)\wedge d\phi_1 \wedge d\phi_2\label{brend} \\
H_3 &=&  {6g_s {\bf A_4} M_{\rm eff}}\Bigg(1+\frac{9g_s N^{\rm eff}_f}{4\pi}~{\rm log}~r+\frac{g_s N^{\rm eff}_f}{2\pi} 
~{\rm log}~{\rm sin}\frac{\theta_1}{2}~
{\rm sin}\frac{\theta_2}{2}\Bigg)\frac{dr}{r}\nonumber \\
&& \wedge \frac{1}{2}\Bigg({\rm sin}~\theta_1~ d\theta_1 \wedge d\phi_1
- {\bf B_4}~{\rm sin}~\theta_2~ d\theta_2 \wedge d\phi_2\Bigg)
+ \frac{3g^2_s M_{\rm eff} N^{\rm eff}_f}{8\pi} {\bf A_5} \Bigg(\frac{dr}{r}\wedge e_\psi -\frac{1}{2}de_\psi \Bigg)\nonumber  \\
&& \hspace*{1.5cm} \wedge \Bigg({\rm cot}~\frac{\theta_2}{2}~d\theta_2 
-{\bf B_5}~{\rm cot}~\frac{\theta_1}{2} ~d\theta_1\Bigg)\nonumber
\end{eqnarray}
where $\widetilde F_3 \equiv F_3 - C_0 H_3$, $C_0$ being the ten dimensional axion and
the so-called asymmetry factors ${\bf A_i}, {\bf B_i}$ are given in eq. (3.83) of \cite{FEP} (see also \cite{sullyf}). 
The axio-dilaton 
and the five-form fluxes are:
\bg\label{dilato}
&&C_0 ~ = ~ {N^{\rm eff}_f \over 4\pi} (\psi - \phi_1 - \phi_2)\nonumber\\
&& e^{-\Phi}~ =~ {1\over g_s} -\frac{N^{\rm eff}_f}{8\pi} ~{\rm log} \left(r^6 + 9a^2 r^4\right) - 
\frac{N^{\rm eff}_f}{2\pi} {\rm log} \left({\rm sin}~{\theta_1\over 2} ~ {\rm sin}~{\theta_2\over 2}\right)\nonumber\\
&& F_5 ~ = ~ {1\over g_s} \left[d^4 x \wedge d h^{-1} + \ast(d^4 x \wedge dh^{-1})\right]
\nd
with $a$ being the resolution parameter of the internal space that depends on the horizon radius $r_h$ as 
$a = a(r_h) + {\cal O}(g_s^2 M_{\rm eff}N^{\rm eff}_f)$. Once we consider the slice:
\bg\label{sol}\theta_1~ = ~ \theta_2 ~ = ~ \pi,~~~~~~\phi_i~ = ~ 0,~~~~~~~\psi~ = ~0 \nd
the background along the slice simplifies quite a bit. To ${\cal O}(g_sN_f)$ the background is:
\bg\label{slicebg}
&& h =\frac{L^4}{r^4}\Bigg[1+\frac{3g_sM_{\rm eff}^2}{2\pi N_{\rm eff}}{\rm log}r\left\{1+\frac{3g_sN^{\rm eff}_f}{2\pi}\left({\rm
log}r+\frac{1}{2}\right)\right\}\Bigg]\nonumber\\
&& H_3 ~ = ~ \widetilde F_3 ~ = ~ C_0 ~ = ~ 0 \nonumber\\
&& e^{-\Phi}~ =~ {1\over g_s} -\frac{N^{\rm eff}_f}{8\pi} ~{\rm log} \left(r^6 + 9a^2 r^4\right)
\nd
alongwith $F_5$ given by \eqref{dilato}. The simplicity of the background is the reason why our analysis of the 
mass and the drag of the quark in \cite{FEP} were 
straightforward enough to see the underlying physics, yet were not afflicted 
by problems like UV divergences of \cite{cschu}\footnote{On the slice \eqref{sol} the pull-backs of the $B$-fields are 
zero. This means that Wilson loops or other equivalent constructions could be carried out without any interference 
from the logarithmic $B$-fields.}. Note that the logarithmic RG flows of the two couplings come from the logarithmic
$B_{NS}$ field, leading to confinement at the far IR (at zero temperature). In the following, to avoid 
clutter, ($N, N_f, M$) would 
denote their effective values.

\subsection{Region 2: Interpolating Region and the Detailed Background}

To attach a UV cap that allows a vanishing beta function we need at least a configuration of vanishing NS three-form.
This cannot be {\it abruptly} attached to Region 1: we need an interpolating region. This region, which we 
will call Region 2,
should have the behavior that at the outermost boundary the three-forms vanish, while solving the equations of 
motion. The innermost boundary of Region 2 $-$ that also forms the outermost boundary of Region 1 $-$ will be determined 
by the scale associated with the mass of the lightest quark, $m_0$, in our system. In terms of  Figure 4, this 
is given by region in the local neighborhood of $r_{\rm min} \equiv m_0 T_0^{-1} + r_h$, where $T_0$ and $r_h$ 
are the string tension and the horizon radius respectively.  
We have already discussed some aspects of this in our previous paper \cite{FEP} when we discussed the issue of 
UV caps. It is now time to spell this in more details. 

The structure of the warp factor should be clear from \cite{FEP}. 
We expect the form to look like \eqref{larger} 
discussed earlier. For our purpose, it would make more sense to rewrite this in such a way that the radial $r$ 
dependence shows up explicitly. 
For this we need to 
first define two functions $f(r)$ and $M(r)$ as (see Figure 5):
\bg\label{mdefo}
f(r) ~ \equiv ~ {e^{\alpha(r-r_0)}\over 1 + e^{\alpha(r - r_0)}}, ~~~~~~~ M(r) ~\equiv~ M [1-f(r)], ~~~~~ \alpha >> 1
\nd  
where the scale $r_0$ will be explained below and $M$ is as before related to the effective number of five-branes (or the 
RR three-form charge). 
Note that for $r << r_0$, $f(r) \approx e^{r-r_0}$, whereas for 
$r > r_0$, $f(r) \approx 1$. Thus for $r$ smaller than the scale $r_0$, $f(r)$ is a very small quantity; whereas for
$r$ bigger than the scale $r_0$, $f(r)$ is identity. In terms of $M(r)$
this means that for $r < r_0$, $M(r) \approx M$ whereas for 
$r > r_0$, $M(r) \to 0$.
This will be useful below. 

Using these functions, we see that the simplest way in which logarithmic behavior 
along the radial direction may go to inverse $r$ behavior, is when the warp factor has the following form:
\bg\label{warpy}
h ~ = ~ { c_0 + c_1 f(r) + c_2f^2(r)\over r^4} \sum_{\alpha} 
~{L_{\alpha} \over r^{\epsilon_{(\alpha)}}}  
\nd
where $c_i$ are constant numbers, and the denominator can be mapped to  
$r_{(\alpha)}$ defined in \eqref{larger} with $\epsilon_{(\alpha)}$ functions of $g_sN_f, M, N$ and the resolution 
parameter $a$. 
$L_{\alpha}$'s are functions of the angular 
coordinates ($\theta_i, \phi_i, \psi$). For other details see \cite{FEP}. 
The warp factor $h$  
has the required logarithmic behavior as long as 
the exponents of $r$ are small and fractional, and indeed switches to the inverse $r$ behavior as soon as 
the exponents become integers.
In \cite{FEP} we gave some examples where the exponents are small and fractional numbers, and
alluded to the case where they become integers\footnote{See the section on holographic 
renormalisability in \cite{FEP}.}. Since
$N_f$ is a delocalised function, this behavior could be naturally realised now and
would eventually give way to the required inverse $r$ behavior of the 
warp factor in Region 3. Its at least clear that such a behavior of the warp factor do solve the background supergravity 
equations of motion near $r = r_{\rm min}$
(see \cite{FEP} for a concrete example, and we will give more details on this below), 
however what we want to know whether such a behavior of the warp factor is generically a solution to 
EOM, or we need to add sources to the theory. It will turn out that we need to add sources at the 
outermost boundary of region 2. Question now is to figure out 
consistently the specific point in the radial direction 
beyond which Region 3 would start. This way we will know exactly {\it where} to add the sources and the AdS cap. 

The demarcation point can be found easily by looking at the behavior of $H_{NS}$ and $H_{RR}$. For this we need to 
use the functions \eqref{mdefo} to write the 
RR three-form. Our ansatze for ${\widetilde F}_3$ then is:
\begin{eqnarray}
&&{\widetilde F}_3  = \left({a}_o - {3 \over 2\pi r^{g_sN_f}} \right)
\sum_\alpha{2M(r)c_\alpha\over r^{\epsilon_{(\alpha)}}} 
\left({\rm sin}~\theta_1~ d\theta_1 \wedge d\phi_1- 
\sum_\alpha{f_\alpha \over r^{\epsilon_{(\alpha)}}}~{\rm sin}~\theta_2~ d\theta_2 \wedge
d\phi_2\right)\nonumber\\
&&~~ \wedge~ {e_\psi\over 2}-\sum_\alpha{3g_s M(r)N_f d_\alpha\over 4\pi r^{\epsilon_{(\alpha)}}}   
~{dr}\wedge e_\psi \wedge \left({\rm cot}~{\theta_2 \over 2}~{\rm sin}~\theta_2 ~d\phi_2 
- \sum_\alpha{g_\alpha \over r^{\epsilon_{(\alpha)}}}~ 
{\rm cot}~{\theta_1 \over 2}~{\rm sin}~\theta_1 ~d\phi_1\right)\nonumber \\
&& -\sum_\alpha{3g_s M(r) N_f e_\alpha\over 8\pi r^{\epsilon_{(\alpha)}}}
~{\rm sin}~\theta_1 ~{\rm sin}~\theta_2 \left({\rm cot}~{\theta_2 \over 2}~d\theta_1 +
\sum_\alpha{h_\alpha \over r^{\epsilon_{(\alpha)}}}~ 
{\rm cot}~{\theta_1 \over 2}~d\theta_2\right)\wedge d\phi_1 \wedge d\phi_2\label{brend}
\end{eqnarray}
where $a_o = 1 + {3\over 2\pi}$ and ($c_\alpha, ..., h_\alpha$) are constants. 
One may also notice three things: first, 
how the internal forms get deformed near the innermost boundary of the region, second, how the function $f(r)$ appears
for all the components, and finally, how
$N_f$ is, as before, not a constant but a delocalised function\footnote{We will soon see that $N_f$ in fact 
is the effective number of seven-branes.}. 
The function $f(r)$ becomes identity for 
$r > r_0$ and therefore ${\widetilde F}_3 \to 0$ for $r > r_0$. For $r < r_0$, the corrections coming from
$f(r)$ is exponentially small. Integrating ${\widetilde F}_3$ over the topologically non-trivial three-cycle:
\bg\label{3cycle} 
{1\over 2}{e_\psi} \wedge
\left({\rm sin}~\theta_1~ d\theta_1 \wedge d\phi_1- 
\sum_\alpha{f_\alpha \over r^{\epsilon_{(\alpha)}}}~{\rm sin}~\theta_2~ d\theta_2 \wedge
d\phi_2\right)
\nd
we find that the number of units of RR flux vary in the following way with respect to the radial coordinate $r$:
\bg\label{fvary}
M_{\rm tot}(r) =  M(r) \left(1 + {3\over 2\pi} - {3 \over 2\pi r^{g_sN_f}} \right)
\sum_\alpha{c_\alpha\over r^{\epsilon_{(\alpha)}}}
\nd
which is perfectly consistent with the RG flow, because for $r < r_0$, and $r \to r e^{-{2\pi\over 3 g_s M}}$,
$M_{\rm tot}$ decreases precisely as $M - N_f$ as the correction factor $e^{r-r_0}$ coming from $f(r)$ is 
negligible. For $r > r_0$, $M_{\rm tot}$ shuts off completely. This also means that below $r_0$, the total colors 
$N$ decrease by $M_{\rm tot}$ exactly as one would have expected for the RG flow with $N_f$ flavors. 
\begin{figure}[htb]\label{ffunction}
		\begin{center}
\includegraphics[height=8cm,width=6cm,angle=-90]{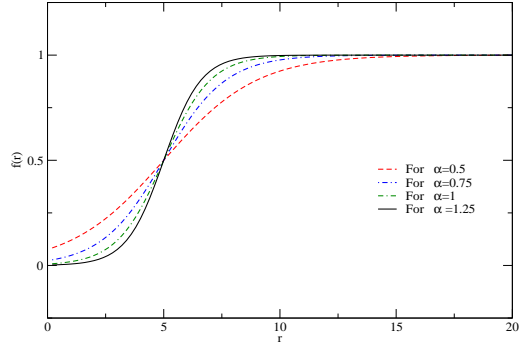}
		\caption{A plot of the $f(r)$ function for $r_0 = 5$ in appropriate units, and
various choices of $\alpha$. Observe that for large 
$\alpha$ the function quickly approaches 1 for $r > r_0$.}
	        \end{center}
		\end{figure}  
Using similar deformed internal forms, one can
also write down the ansatze for the NS three-form. This is given as:
\begin{eqnarray}
&&H_3 =  \sum_\alpha {6g_s M(r) k_\alpha \over r^{\epsilon_{(\alpha)}}}\Bigg[1+\frac{1}{2\pi} - 
\frac{\left({\rm cosec}~\frac{\theta_1}{2}~{\rm cosec}~\frac{\theta_2}{2}\right)^{g_sN_f}}{2\pi r^{{9g_sN_f\over 2}}}
\Bigg]~ dr \nonumber\\
&&\wedge \frac{1}{2}\Bigg({\rm sin}~\theta_1~ d\theta_1 \wedge d\phi_1
-\sum_\alpha{p_\alpha \over r^{\epsilon_{(\alpha)}}} ~{\rm sin}~\theta_2~ d\theta_2 \wedge d\phi_2\Bigg)
+\sum_\alpha \frac{3g^2_s M(r) N_f l_\alpha}{8\pi r^{\epsilon_{(\alpha)}}}
\Bigg(\frac{dr}{r}\wedge e_\psi -\frac{1}{2}de_\psi \Bigg)\nonumber\\
&& \wedge \Bigg({\rm cot}~\frac{\theta_2}{2}~d\theta_2 
-\sum_\alpha{q_\alpha \over r^{\epsilon_{(\alpha)}}}~{\rm cot}~\frac{\theta_1}{2} 
~d\theta_1\Bigg) + 
g_s {dM(r) \over dr}
\left(b_1(r)\cot\frac{\theta_1}{2}\,d\theta_1+b_2(r)\cot\frac{\theta_2}{2}\,d\theta_2\right)\nonumber\\
&&\wedge e_\psi
\wedge dr
+{3g_s\over 4\pi} {dM(r) \over dr}\left[\left(1+g_sN_f -{1\over r^{2g_sN_f}} + {9a^2g_sN_f\over r^2}\right)
\log\left(\sin\frac{\theta_1}{2}\sin\frac{\theta_2}{2}\right) + b_3(r)\right]\nonumber\\
&& \sin\theta_1\,d\theta_1\wedge d\phi_1\wedge dr
-{g_s\over 12\pi}{dM(r) \over dr} \Bigg(2 -{36a^2g_sN_f\over r^2} + 9g_sN_f -{1\over r^{16g_sN_f}} - 
{1\over r^{2g_sN_f}} + {9a^2g_sN_f\over r^2}\Bigg)\nonumber\\ 
&& ~~~~~~~~~~~~~~~~ \sin\theta_2\,d\theta_2\wedge d\phi_2\wedge dr - 
{g_sb_4(r)\over 12\pi}{dM(r) \over dr}~ \sin\theta_2\,d\theta_2\wedge d\phi_2\wedge dr
\label{brend2}
\end{eqnarray} 
with ($k_\alpha, ..., q_\alpha$) being constants and $b_n = \sum_m {a_{nm}\over r^{m + \widetilde{\epsilon}_m}}$ 
where $a_{nm} \equiv a_{nm}(a^2, g_sN_f)$ and $\widetilde{\epsilon}_m \equiv \widetilde{\epsilon}_m(g_sN_f)$. 
The way we constructed the three-forms imply that $H_3$ is closed. In fact the 
${\cal O}(\partial f)$ terms that we added to \eqref{brend2} ensures that. However $F_3$ is not closed. 
We can use the non-closure of $F_3$ to analyse {\it sources} that we need to add for consistency. 
These sources should in general be ($p, q$) five-branes, with ($p, q$) negative,
 so that they could influence both the 
three-forms and 
since the ISD property of the three-forms is satisfied near $r = r_{\rm min}$ the sources should be 
close to the other boundary. A simplest choice could probably just be anti five-branes because adding anti D5-branes
would change ${\widetilde F}_3$, and to preserve the ISD condition, $H_3$ would have to change accordingly. 
Furthermore, as we mentioned before,
as $r \to r_0$, both $H_3 = {\widetilde F}_3 \to 0$. Therefore 
$r = r_0$ is where Region 2 ends and Region 3 begins, and we can put the sources there. They could be oriented along
the spacetime directions, located around the local neighborhood of $r = r_0$ and wrap the internal two-sphere 
($\theta_1, \phi_1$) so that they are parallel to the seven-branes.
However, putting in anti D5-branes near $r= r_0$ would imply non-trivial forces between the five-branes 
and seven-branes as well as five-branes themselves. Therefore if we keep, in general the ($p, q$) 
the five-branes close to say one
of the seven-brane then they could get {\it dissolved} in the seven-brane as electric and magnetic
gauge fluxes $\ast F^{(1)}$ and $F^{(1)}$ respectively. Thus the 
seven-brane soaks in the five-brane charges, which in turn would mean that ${\widetilde F}_3$ in \eqref{brend} 
and $H_3$ in \eqref{brend2} will satisfy the following EOMs:
\bg\label{eombrend}
d{\widetilde F}_3~&=&~ F^{(1)} \wedge \Delta_2(z) - d\left({\bf Re}~\tau\right) \wedge H_3\nonumber\\
d\ast H_3 ~&=&~ \ast F^{(1)} \wedge \Delta_2(z) - d(C_4 \wedge F_3)
\nd
where the tension of the seven-brane is absorbed in $\Delta_2(z)$, which 
is the term that measures the delocalisation of the seven branes (for localised 
seven branes this would be copies of
the two-dimensonal delta functions) and $\tau$ is the axio-dilaton that we will determine below. In addition to 
that $d\ast F_3$ will satisfy its usual EOM. 
For all the analysis in this paper we will also assume:
\bg\label{regide}
\vert r_0 - r_{\rm min}\vert ~\le~ a_1, ~~~~ \vert r_{\rm min} - r_h \vert ~\le ~a_2, ~~~~ {\rm Region}~3 ~>>~ a_1 + a_2
\nd
to be our approximation. This way, as we said before, Region 3 will dominate our calculations. 

However the above set of equations \eqref{eombrend} is still not the full story. Due to the anti GSO projections 
between anti-D5 and D7-brane, there should be tachyon between them. It turns out that the tachyon can be 
removed (or made massless) by switching on additional electric and magnetic fluxes on D7
along, say, ($r, \psi$)
directions! This would at least kill the instability due to the tachyon, although susy 
may not be restored. For details on the precise 
mechanism, the readers may refer to \cite{susyrest}. But switching on gauge fluxes on D7 would generate 
extra D5 charges and switching on gauge fluxes on anti-D5s will generate extra D3 charges. This is one reason 
why we write ($N, N_f, M$) as effective charges. This way a stable system of anti-D5s and D7 could be constructed.  


To complete the rest of the 
story we need the axio-dilaton $\tau$ and the five-form. The five-form is easy to determine
from the warp factor $h$ \eqref{warpy} using \eqref{dilato}. The total five-form charge should have contribution 
from the gauge fluxes also, which in turn would effect the warp factor.  
For regions close to $r_{\rm min}$ it is clear 
that $\tau$ goes as $z^{-g_sN_f}$ where $z$ is the embedding \eqref{embedding}. More generically and for the whole of 
Region 2, looking at the warp factor and the three-form fluxes, we expect the axio-dilaton to go 
as\footnote{One may use this value of axio-dilaton and the three-form NS fluxes \eqref{brend2} to determine the 
beta function from the relations \eqref{maps}. To lowest order in $g_sN_f$ we will reproduce the SV beta function 
\eqref{svbeta} as expected. Notice that for $r > r_0$ the beta function {\it does not} vanish and both the gauge 
groups flow at the same rate. This will be crucial for our discussion in the following subsection.}:
\bg\label{axdilato}
\tau ~=~ [b_0 + b_1 f(r)]\sum_\alpha {C_\alpha\over r^{\epsilon_{(\alpha)}}}
\nd
where $b_i$ are constants and $C_\alpha$ are functions of the internal coordinates and are complex. These
$C_\alpha$ and the constants $b_i$ are determined from the dilaton equation of motion \cite{drs, gkp}:
\bg\label{dieq}
{\widetilde\nabla}^2~\tau = {{\widetilde\nabla}\tau\cdot {\widetilde\nabla}\tau\over i{\rm Im}~\tau} - 
{4\kappa_{10}^2 ({\rm Im}~\tau)^2\over \sqrt{-g}} {\delta S_{\rm D7}\over \delta\bar\tau} + (p,q)~ 
{\rm sources}
\nd
where tilde denote the unwarped internal metric $g_{mn}$, and $S_{\rm D7}$ is the action for the {\it delocalised} 
seven branes. The $f(r)$ term in the axio-dilaton come from the ($p, q$) sources that are absorbed as gauge fluxes 
on the seven-branes\footnote{The $r^{-\epsilon_{(\alpha)}}$ behavior stems from additional anti seven-branes that we 
need to add to the existing system to allow for the required UV behavior from the F-theory completion. 
The full picture will become clearer in the next sub-section when we 
analyse the system in Region 3.}.
Because of this behavior of axio-dilaton we don't expect the unwarped metric to remain Ricci-flat 
to the lowest order in $g_sN_f$. The Ricci tensor becomes:
\bg\label{rten}
{\widetilde{\cal R}}_{mn} = \kappa^2_{10} {\partial_{(m}\partial_{n)}\tau\over 4({\rm Im}~\tau)^2} + 
\kappa_{10}^2 \left({\widetilde T}^{\rm D7}_{mn} - {1\over 8}{\widetilde g}_{mn}{\widetilde T}^{\rm D7}\right)
+ \kappa_{10}^2 \left({\widetilde T}^{(p,q)5-{\rm brane}}_{mn} - 
{1\over 4}{\widetilde g}_{mn}{\widetilde T}^{(p,q)5-{\rm brane}}\right)\nonumber\\
\nd 
where we see that ${\widetilde{\cal R}}_{rr}$ picks up terms proportional to $\epsilon^2_{(\alpha)}$ 
and derivatives of $f(r), N_f(r)$, implying that to zeroth order in $g_sN_f$ the interpolating region 
may not remain Ricci-flat. However since the coefficients 
are small, the deviation from Ricci-flatness is consequently small.
In this paper we will not give the explicit form for 
$C_\alpha, L_\alpha$ etc but it should be clear from our above discussions that EOMs are easily satisfied.
The one last thing to check 
would be the equation for the warp factor. This is given by the five-form equation of motion:
\bg\label{5form}
d\ast d h^{-1} = H_3 \wedge {\widetilde F}_3 + \kappa_{10}^2 ~{\rm tr} 
\left(F^{(1)}\wedge F^{(1)} - {\cal R}\wedge {\cal R}\right)
\Delta_2(z) + \kappa_{10}^2 ~{\rm tr}~F^{(2)} {\widetilde \Delta}_4({\cal S})\nonumber\\
\nd
where $F^{(1)}$ is the seven-brane gauge fields that we discussed earlier, 
$F^{(2)}$ is the ($p, q$) five-brane gauge fields required for the proper interpretation of the colors in the 
gauge theory side\footnote{In fact one should view the gauge fluxes on the seven-branes and the five-branes as the 
total gauge fluxes that are needed to stabilise the system. We will see in the next subsection that the full stabisation
would require additional fluxes, but the structure would remain the same.},
${\cal R}$ is the pull-back of the Riemann two-form,
and ${\widetilde \Delta}_4({\cal S})$ is the term that 
measures the delocalisation of the dissolved ($p, q$) five-branes over the space ${\cal S}$ embedded in the seven-brane
(again for localised five-branes there would be copies of four-dimensional
delta functions). The $H_3 \wedge {\widetilde F}_3$ term in \eqref{5form} is proportional to ${M^2(r)\over 
r^{2\epsilon_{(\alpha)}}}$. This is precisely the form for the warp factor ansatze \eqref{warpy} with the 
$f^2(r)$ term there accounting for the $M^2(r)$ term above. This way
with the warp factor \eqref{warpy} and the 
three-forms \eqref{brend} and \eqref{brend2} we can satisfy \eqref{5form} by switching on small gauge fluxes on the
seven-branes and five-branes. 

Therefore combining 
\eqref{warpy}, \eqref{brend}, \eqref{brend2}, \eqref{axdilato} and the five-form, we can pretty much determine the 
supergravity background for the interpolating region $r_{\rm min} < r \le r_0$. At the outermost boundary of Region 2
we therefore
only have the metric and the axio-dilaton. Both the three-forms exponentially decay away fast, 
giving us a way to attach an
AdS cap there.    

\subsection{Region 3: Seven Branes, F-Theory and UV Completions} 

The interpolating region, Region 2, that we derived above can be interpreted alternatively as the {\it deformation} 
of the neighboring geometry once we attach an AdS cap to the OKS-BH geometry. The OKS-BH geometry is the range 
$r_h \le r \le r_{\rm min}$ and the AdS cap is the range $r > r_0$. The geometry in the 
range $r_{\rm min} \le r \le r_0$ is the deformation. Such deformations should be expected for all other UV caps 
advocated in \cite{FEP}. In this section we will complete the rest of the picture by elucidating the background 
from $r > r_0$ in the AdS cap. But before that let us give a brief gauge theory interpretation of 
background\footnote{The discussion in the following paragraph is motivated by a correspondence that we had with 
Peter Ouyang. We thank him for his comments.}. 

For the UV region $r > r_0$ we expect the dual gauge theory to be $SU(N + M) \times SU(N + M)$ with fundamental 
flavors coming from the seven-branes. This is because addition of ($p, q$) branes at the junction, or more appropriately 
anti five-branes at the junction with gauge fluxes on its world-volume, tell us that the number of three-branes
degrees of freedom are $N + M$, with the $M$ factor coming from five-branes anti-five-branes pairs. 
Furthermore, the $SU(N + M) \times
SU(N + M)$ gauge theory will tell us that the gravity dual is approximately AdS, but has RG flows because of the 
fundamental flavors (This RG flow is the remnant of the flow that we saw in the 
previous subsection. We will determine this in more details below). 
At the scale $r = r_0$ we expect one of the gauge group 
to be Higgsed, so that we are left with $SU(N + M) \times SU(N)$. Now both the gauge fields flow at different rates 
and give rise to the cascade that is slowed down  by the $N_f$ flavors. In the end, at far IR, we expect 
confinement at zero temperature.

The few tests that we did above, namely, (a) the flow of $N$ and $M$ colors, 
(b) the RG flows, (c) the decay of the 
three-forms, and (d) the behavior of the dual gravity background, all point to the gauge theory interpretation that we
gave above. What we haven't been able to demonstrate is the precise Higgsing that takes us to the cascading 
picture. From the gravity side its clear how this could be interpreted. From the gauge theory side it would be 
interesting to demonstrate this. 

Coming back to the analysis of Region 3, we see that 
in the region $r > r_0$ we do not expect three-forms but we do expect non-zero axio-dilaton. These non-zero axio-dilaton
come from the rest of the seven branes. As mentioned in \cite{FEP} the complete set of seven-branes should be determined
from the F-theory picture \cite{vafaF} to capture the full non-perturbative corrections. This is now subtle because the 
seven-branes are embedded non-trivially here (see \eqref{embedding}). A two-dimensional base, parametrised by a 
complex coordinate $z$, 
on which we 
can have a torus fibration:
\bg\label{torus}
y^2 = x^3 + x F(z) + G(z)
\nd
can be identified with the $z$ coordinate of \eqref{embedding}. This way vanishing discriminant 
$\Delta$ of \eqref{torus} i.e $\Delta \equiv 4F^3 + 27G^2 = 0$, will specify the positions of the seven-branes exactly as
\eqref{embedding}. Here we have taken $F(z)$ as a degree eight polynomial in $z$ and $G(z)$ as a degree 12 polynomial 
in $z$. The delocalisation ${\widetilde N}_f(r, \theta_2, \phi_2)$ should be thought of somewhat 
as the distribution of bunches of seven branes along ($\theta_2, \phi_2$) directions with varying {\it sizes} along 
the radial $r$ direction such that \eqref{nfconstraint} is maintained with the deviation
$\delta \equiv {\hat r} - r_0$ a finite but not very large number.

As is well known, embedding of seven-branes in F-theory also tells us that we can have $SL(2, {\bf Z})$ jumps of 
the axio-dilaton. We can define the axio-dilaton $\tau \equiv C_0 + i e^{-\phi}$ as the modular parameter of a torus 
${\bf T}^2$ fibered over the base parametrised by the coordinate $z$. The holomorphic map\footnote{Holomorphic in 
$\tau$, the modular parameter.} 
from the fundamental domain
of the torus to the complex plane is given by the famous $j$-function:
\bg \label{axdil1}
j(\tau) ~\equiv ~ \frac{\left[\Theta_1^8(\tau)+\Theta_2^8(\tau)+\Theta_3^8(\tau)\right]^3}{\eta^{24}(\tau)} ~= ~
\frac{4(24{F}(z))^3}{27{G}^2(z)+4{F}^3(z)}
\nd
where $\Theta_i, i = 1, 2, 3$ are the well known Jacobi Theta-functions and $\eta$ is the Dedekind 
$\eta$-function:
\bg\label{dedekind}
\eta(\tau) ~=~ q^{1\over 24}\prod_n (1 - q^n),~~~~~~~~~ q ~= ~ e^{2\pi i \tau}
\nd
For our purpose, we can write the discriminant $\Delta(z)$ and the polynomial $F(z)$ generically as:
\bg\label{delF}
\Delta(z) ~=~ 4F^3 + 27G^2 ~=~ a \prod_{j =1}^{24} (z - {\widetilde z}_j), ~~~~~~ F(z) ~=~ b \prod_{i = 1}^8 (z - z_i)
\nd
so that when we have weak type IIB coupling i.e $\tau = C_0 + i\infty$, $j(\tau) \approx e^{-2\pi i \tau}$ and 
using \eqref{axdil1}
the modular parameter can be mapped to the embedding coordinate $z$ as:
\bg\label{modmap}
\tau ~ &=& ~ {i\over g_s} ~+~
{i\over 2\pi} ~{\rm log}~(55926 ab^{-1}) - {i\over 2\pi} \sum_{n = 1}^\infty \left[{1\over nz^n} 
\left(\sum_{i=1}^8 3 z_i^n -  \sum_{j=1}^{24} {\widetilde z}_j^n\right)\right] \nonumber\\
&=&~ \sum_{n = 0}^\infty {{\cal C}_n + i{\cal D}_n\over {\widetilde r}^n}
\nd
where ${\cal C}_n \equiv {\cal C}_n(\theta_i, \phi_i, \psi)$ and ${\cal D}_n \equiv {\cal D}_n(\theta_i, \phi_i, \psi)$
are real functions and ${\widetilde r} = r^{3/2}$. To avoid cluttering
of formulae, we will use $r$ instead of ${\widetilde r}$ henceforth unless mentioned otherwise. So the 
coordinate $r$ will parametrise Region 3, and $\tau = \sum {{\cal C}_n + i{\cal D}_n\over r^n}$.

The above computation was done
assuming that
$z > (z_i, {\widetilde z}_j)$, which at this stage can be guaranteed if we take $\theta_{1,2}$ small. 
This gives rise to special
set of configurations of seven-branes where they are distributed along other angular directions. 
However one might get a little worried if there exists some ${\widetilde z}_j \equiv {\widetilde z}_o$ related 
to the {\it farthest} seven-brane(s) where the above approximation fails to hold. This can potentially happen 
when we try to compute the mass of the heaviest quark in our theory. The question is whether we can still use the 
$\tau$ derived in \eqref{modmap}, or we need to modify the whole picture. 

Before we go into answering this question, the choice of $z$ bigger than ($z_i, {\widetilde z}_j$) 
already needs more
convincing
elaboration because allowing $\theta_{1,2}$ small is a rather naive argument. The situation at hand is more subtle 
than that and, as we will argue below, the picture that we have right now is incomplete.
 
To get the full picture, observe first that 
$z$ being given by our embedding equation \eqref{embedding}, means that if we want to 
be in Region 3, we need to specify the condition $r > r_0$ in the defination of $z$. This way a given $z$ will 
{always} imply points in Region 3 for varying choices of the angular coordinates ($\theta_i, \phi_i, \psi$). 
However similar argument cannot be given for any choices of 
($z_i, {\widetilde z}_j$). A particular choice of ($z_i, {\widetilde z}_j$) may imply very large $r$ with small 
angular choices or small $r$ with large angular choices. Thus analysing the system 
only in terms of the $r$ coordinate is tricky. In terms of the full complex coordinates, 
$z > (z_i, {\widetilde z}_j)$ would mean that we are always looking at 
points away from the surfaces given by $z = z_i$ and $z = {\widetilde z}_j$. 

What happens when we touch the $z = z_i$ surfaces? 
For these cases $F(z_i) \to 0$ and therefore we are no longer
in the weak coupling regime. For all $F(z_i) = 0$ imply $j(\tau) \to 0$ which in turn means 
$\tau = {\rm exp}~(i\pi/3)$ on these surfaces. 
These are the constant coupling regimes of \cite{DM} where the string couplings on these surfaces are 
 {\it not} weak. On the other hand, 
near any one of the seven-branes $z = {\widetilde z}_j$ we are in the 
weak coupling regimes and so \eqref{modmap} will imply 
\bg\label{taulog}
\tau(z) = {1\over 2\pi i} ~{\rm log}~(z - {\widetilde z}_j) ~\to~i\infty
\nd
which of course is 
expected but nevertheless problematic for us. This is because we need logarithmic behavior 
of axio-dilaton in Region 2, but not in Region 3. For a good UV behavior, we need axio-dilaton to behave like 
\eqref{modmap} everywhere in Region 3. 

In addition to that there is also the issue of the heaviest 
quarks creating additonal log divergences that we mentioned earlier. These seven branes are located at 
$z = {\widetilde z}_j \equiv {\widetilde z}_o$, and therefore
if we can make the axio-dilaton independent of the coordinates
 ${\widetilde z}_o$ then at least we won't get any divergences from these seven-branes. 
It turns out that there are configurations (or rearrangements) of seven-brane(s) that allow us to do exactly 
that. To see one such configuration, let us define $F(z), G(z)$ and $\Delta(z)$ in 
\eqref{delF} in the following way:
\bg\label{delFnow}
&&F(z) ~ = ~ (z - {\widetilde z}_o)\prod_{i = 1}^7 (z - z_i), ~~~~~~~~~
G(z) ~=~ (z- {\widetilde z}_o)^2 \prod_{i =1}^{10} 
(z - {\hat z}_i)\nonumber\\ 
&&\Delta(z) ~ = ~ (z - {\widetilde z}_o)^3 \prod_{j = 1}^{21} (z - {\widetilde z}_j) 
\nd
which means that we are stacking a bunch of {\it three} seven-branes at the point $z = {\widetilde z}_o$, and 
\bg\label{deldefn}
\prod_{j = 1}^{21} (z - {\widetilde z}_j) ~ \equiv ~ 
4\prod_{i = 1}^7 (z - z_i)^3 ~ + ~ 27 (z - {\widetilde z}_o) \prod_{i = 1}^{10} 
(z - {\hat z}_i)^2
\nd
implying that the axio-dilaton $\tau$ becomes independent of ${\widetilde z}_o$ and behaves
exactly as in \eqref{modmap} with ($i, j$) in \eqref{modmap} varying 
upto (7, 21) respectively. 

The situation is now getting better. We have managed to control a subset of log divergences. 
To get rid of the other set of log divergences that appear on the remaining 
twenty-one surfaces, one possible way would be  
to modify the embedding \eqref{embedding}. Recall that our configuration is 
non-supersymmetric and therefore we are not required to use the embedding \eqref{embedding}. In fact a change in the 
embedding equation will also explain the axio-dilaton choice \eqref{axdilato} of Region 2. 
To change the embedding equation \eqref{embedding} we will 
use similar trick that we used to kill off the three-form fluxes, namely, attach anti-branes. These
anti seven-branes\footnote{They involve both local and non-local anti seven-branes.} 
are embedded via the following equation:
\bg \label{ABembedding}
r^{3/2} e^{i(\psi-\phi_1-\phi_2)}{\rm sin}~\frac{\theta_1}{2}~{\rm sin}~\frac{\theta_2}{2}~=~r_0 e^{i\Theta}
\nd
where $\Theta$ is some angular parameter, and could vary for different anti seven-branes. The above embedding 
will imply that their overlaps with 
the corresponding seven-branes are only partial\footnote{For example if we have a seven-brane at $z = {\widetilde z}_1$ 
such that lowest point of the seven brane is $r = \vert {\widetilde z}_1\vert^{2/3} < r_o$, 
then the corresponding anti-brane 
has only partial overlap with this.}. And since we require
$$ \vert{\widetilde z}_j\vert^{2/3} ~ < ~ r_o$$
it will appear effectively that we can only have seven-branes in 
Regions 1 and 2, and {\it bound} states of seven-branes and anti seven-branes in Region 3.\footnote{Of course this 
effective descrition is only in terms of the axio-dilaton charges. In terms of the embedding equation for the 
seven-branes \eqref{embedding} this would imply that we can define 
$z$ with $r > r_o$ and ${\widetilde z}_j$ with $r < r_o$.}
This way 
the axio-dilaton in Region 3 will indeed behave as \eqref{modmap} for all $z$ 
(except for the above mentioned seven points).  

There are two loose ends that we need to tie up to complete this side of the story. The first one is the issue of 
Gauss' law, or more appropriately, charge conservation. The original configuration of 
24 seven branes had zero global charge, but now with  
the addition of anti seven-branes charge conservation seems to be problematic. There are a few ways to resolve this issue.
First, we can asume that that branes wrap topologically trivial cycles, much like the ones of \cite{ouyang}. Then 
charge conservation is automatic. The second alternative is to isolate six seven-branes using some 
appropriate $F$ and $G$ functions, so that they are charge neutral. This is of course one part of the constant 
coupling scenario of \cite{senF}. Now if we make the ($\theta_2, \phi_2$) directions non-compact then we can 
put in a configuration of 18 seven-branes and anti seven-branes pairs together using the embeddings \eqref{embedding} and 
\eqref{ABembedding} respectively. The system would look effectively like what we discussed above. Since the whole 
system is now charge neutral, compactification shouldn't be an issue here. 

The second loose end is the issue of tachyons between the seven-brane and anti seven-brane pairs. Again, as for the 
anti-D5 branes and D7-brane case \cite{susyrest}, 
switching on appropriate electric and magnetic fluxes will make the tachyon massless! 
Therefore the system will be stable and would behave exactly as we wanted, namely, the axio-dilaton will not have the 
log divergences over any slices in Region 3.  

This behavior of axio-dilaton justifies the $r^{-\epsilon_{(\alpha)}}$ in \eqref{axdilato} 
in Region 2. So the 
full picture would be a set of seven-branes with electric and magnetic fluxes embedded via \eqref{embedding}
and another set of anti seven-branes embedded via \eqref{ABembedding} lying completely in Region 3.
 
Thus in Region 3 both the three-forms vanish and therefore $g_1 = g_2 = g_{\rm YM}$ with $g_1, g_2$ being the 
couplings for $SU(N+M), SU(N+M)$. From \eqref{maps} we can compute the $\beta$-function for $g_{\rm YM}$ as:
\bg\label{betym}
\beta(g_{\rm YM}) ~\equiv~ {\partial g_{\rm YM}\over \partial {\rm log}~\Lambda} 
~ = ~ {g^3_{\rm YM}\over 16 \pi}~ \sum_{n = 1}^\infty ~ 
{n{\cal D}_n \over \Lambda^n}
\nd
where $\Lambda$ is the usual RG scale related to the radial coordinate in the supergravity approximation. For
$\Lambda \to \infty$, $\beta(g_{\rm YM}) \to 0$ implying a conformal theory in the far UV. We can fix the 
't Hooft coupling to 
be strong to allow for the supergravity approximation to hold consistently at least for all points away from the 
$z = z_i, i= 1,..., 7$ surfaces.

Existence of axio-dilaton $\tau$ of the form \eqref{modmap} and the seven-brane sources will tell us, from \eqref{rten}, 
that the unwarped metric may not remain Ricci flat. For example it is easy to see that 
\bg\label{rrr}
\widetilde{\cal R}_{rr} = {{\cal A}_{\cal D}\over r^2{\cal D}_0^2} \sum_{n,m =1}^{\infty} nm 
{({\cal C}_n + i{\cal D}_n)({\cal C}_m - i{\cal D}_m)\over r^{n+m}} 
+ {\cal O}\left({1\over r^n}\right)
\nd
where the last term should come from the seven-brane sources and, because of these sources, we don't expect 
$\widetilde{\cal R}_{rr}$ to vanish to lowest order in $g_sN_f$.\footnote{Although, as discussed before, the deviation
from Ricci flatness will be very small.}
 The term ${\cal A}_{\cal D}$ is given by the 
following infinite series:
\bg\label{adddef}
{\cal A}_{\cal D} ~ = ~ 1-\sum_{k,l=1}^\infty {{\cal D}_k {\cal D}_l
{\cal D}_0^{-2}\over r^{k+l}} 
+ \sum_{k,l,p,q=1}^\infty {{\cal D}_k{\cal D}_l{\cal D}_p{\cal D}_q {\cal D}_0^{-2}\over r^{k+l+p+q}} + ...
\nd
Similarly one can show that 
\bg\label{rab}
\widetilde{\cal R}_{ab} = {{\cal A}_{\cal D}\over {\cal D}_0^2}
\sum_{n,m=0}^\infty {(\partial_a{\cal C}_n + i\partial_a{\cal D}_n)
(\partial_b{\cal C}_m - i\partial_b{\cal D}_m)\over r^{n+m}}
+ {\cal O}\left({1\over r^n}\right)
\nd
for ($a, b$) $\ne r$. For $\widetilde{\cal R}_{rb}$ similar inverse $r$ dependence can be worked out. In the far UV 
we expect the 
unwarped curvatures should be equal to the AdS curvatures. The warp factor $h$ on the other hand can be determined 
from the following variant of \eqref{5form}:
\bg\label{wfac}
d\ast d h^{-1} = \kappa_{10}^2 ~{\rm tr} 
\left(F^{(1)}\wedge F^{(1)} - {\cal R}\wedge {\cal R}\right)
\Delta_2(z) + ...
\nd
because we expect no non-zero three-forms in Region 3. The dotted terms are the non-abelian corrections from the 
seven-branes. As $r$ is increased i.e $r >> r_0$, 
we expect $F^{(1)}$ to fall-off (recall that they appear from the anti (1,1) five-branes located in the neighborhood of 
$r = r_0$) and therefore can be absorbed in ${\cal R}$. 
Once we embed the seven-brane gauge connection in some part of spin-connection, we expect 
\bg\label{boxh}
\square ~h^{-1} ~ = ~ {\cal O}\left({1\over r^n}\right)
\nd
from the non-abelian corrections via pull-backs. Solving this will reproduce the generic form for $h$:
\bg\label{heaft}
h ~= ~ \frac{L^4}{r^4}\left[1+\sum_{i=1}^\infty\frac{a_i(\psi,\theta_i, \phi_i)}{r^i}\right]
\nd
with a constant $L^4$ and $a_i$'s are suppressed by powers of $g_sN_f$. 
More details on this is given in the {\bf Appendix A} and {\bf B}. At far UV we recover the AdS picture 
implying a strongly coupled conformal behavior in the dual gauge theory. 

From the above discussion we can conclude that the warp factor 
and the axio-dilaton will have the inverse $r$ behavior. We will use this background to do the Wilson loop computation
in the next section.  

\section{Heavy Quark Potential from Gravity}

Before we go into the actual computation of the Wilson loop, let us point out some generic standard arguments that 
map the Wilson loop computation to the string action and then to the quark anti-quark potential. 

Consider the Wilson loop of a rectangular path ${\cal C} $ with spacelike width $d$ and timelike length $T$.
The timelike
paths can be thought of as world lines of pair of quarks $Q\bar{Q}$ separated by a spatial distance $d$. 
Studying the expectation value of the Wilson loop in the
limit $T\rightarrow \infty$, one can show that it behaves as 
\bg \label{WL-1}
\langle W({\cal C})\rangle ~\sim~ {\rm exp}(-T E_{Q\bar{Q}})
\nd   
where $E_{Q\bar{Q}}$ is the energy of the $Q\bar{Q}$ pair which we can identify with their 
potential energy $V_{Q\bar{Q}}(d)$ as the quarks are static. At this
point we can use the principle of holography \cite{Mal-1} \cite{Witt-1} \cite{Mal-2} 
and identify the expectation value of the Wilson loop with 
the exponential of the {\it renormalised} Nambu-Goto action,
\bg \label{Holo-1}
 \langle W({\cal C})\rangle ~\sim ~ {\rm  exp}(-S^{\rm ren}_{{\rm NG}}) 
\nd
with the understanding that ${\cal C}$ is now the boundary of string world sheet. 
Note that we are computing Wilson loop of gauge theory living on
flat four dimensional space-time $x^{0, 1, 2, 3}$. Whereas the string worldsheet is embedded in curved five-dimensional 
manifold with coordinates
$x^{0, 1, 2, 3}$ and $r$. We will identify the five-dimensional manifold with Region 3 that we discussed above. 
 
To be consistent with the recipe in \cite{Witt-1}, we need to make sure that the induced four dimensional metric 
at the boundary
of the string world sheet ${\cal C}$ is flat. For an AdS space, this is guranteed as long as the world sheet ends on 
boundary of AdS space
where the induced four dimensional metric can indeed be written as $\eta_{\mu\nu}$. 
Using the geometry  constructed in the 
previous section for Region 3,
we see that the metric is asymptotically AdS and therefore induces a flat Minkowski metric at the boundary via:
\bg\label{induce}
\lim_{u \rightarrow 0}~ u^2 g_{\mu\nu}~ = ~ \eta_{\mu\nu}
\nd
where $u = r^{-1}$ and $g_{\mu\nu}$ is the full metric (including the warp factor) in Region 3. 
Thus we can
make the identification (\ref{Holo-1}). Once this subtlety is resolved, comparing (\ref{WL-1}) and (\ref{Holo-1}) 
we can read off the 
potential  
\bg \label{Vqq}
V_{Q\bar{Q}}~ = ~ \lim_{T \to \infty} \frac{S^{\rm ren}_{{\rm NG}}}{T}
\nd      
Thus knowing the renormalised 
string world sheet action, we can compute $V_{Q\bar{Q}}$ for a strongly coupled gauge theory.

The above discussion was all for gauge theory at zero temperature. What happens when we allow non-zero 
temperatures? Does the above identification \eqref{Vqq} between the quark anti-quark potential and the 
renormalised Nambu-Goto action go through again? 

The answer is yes, but the derivation is a little more subtle than what we presented for the zero temperature case. 
At high temperatures and density we expect the medium effects to {\it screen} the interaction between the heavy quark 
and anti-quark pairs. The resulting effective potential between the quark anti-quark pairs separated by a distance $d$
at temperature ${\cal T}$
can then be expressed 
succinctly in terms of the free energy $F(d, {\cal T})$, which generically takes the following form:
\bg\label{freeenergy}
F(d, {\cal T}) = \sigma d ~f_s(d, {\cal T}) - {\alpha\over d} f_c(d, {\cal T})
\nd
where $\sigma$ is the string tension, $\alpha$ is the gauge coupling and $f_c$ and $f_s$ are the screening 
functions\footnote{We expect the screening functions $f_s, f_c$ to equal identity when the temperature goes to 
zero. This gives the zero temperature Cornell potential.} 
(see for example \cite{karsch} and references therein). For the quark and the anti-quark pair kept at 
$+{d\over 2}$ and $-{d\over 2}$ we expect the Wilson lines $W\left(\pm {d\over 2}\right)$ to be related to the free 
energy via:
\bg\label{wlfe}
{\rm exp}\left[-{F(d, {\cal T})\over {\cal T}}\right] ~ = ~ 
{\langle W^\dagger\left(+{d\over 2}\right) W\left(- {d\over 2}\right)\rangle \over 
\langle W^\dagger\left(+{d\over 2}\right)\rangle \langle W\left(-{d\over 2}\right)\rangle}
\nd
In terms of Wilson loop, the free energy \eqref{freeenergy} is now related to the renormalised Nambu-Goto 
action for the string on a background with a black-hole\footnote{There is a big literature on the 
subject where quark anti-quark potential has been computed using various different approaches like 
pNRQCD \cite{brambilla}, hard wall AdS/CFT \cite{polstrass, boschi} and other 
techniques \cite{reyyee, cotrone2}. Its reassuring
to note that the results that we get using our newly constructed background matches very well with the results
presented in the above references. This tells us that despite the large $N$ nature there is an underlying 
universal behavior of the confining potential.}. One may also note that the theory we get is a four-dimensional theory 
{\it compactified} on a circle in Euclideanised version and not a three-dimensional theory.   

\subsection{Computing the Nambu-Goto Action: Zero Temperature} 

Our first attempt to compute the NG action would be to consider the zero temperature case. This means that we make
the black-hole factors $g_i$ in \eqref{bhmet1} to be identity. The string configuration that we will take to 
do the required computation is 
given below in Figure 6. Note that we have configured our geometry such that the string is exclusively in 
Region 3. We will provide a stronger motivation for this soon. For the time being observe that the configuration in 
Figure 6 has one distinct advantage over all other configurations studied in the literature, namely, that 
because of the absence of three-forms in Region 3 we will not have the UV divergence of the Wilson loop 
attributed to the logarithmically varying $B$ field \cite{cschu}. 
In fact even if the string enters Regions 2 and 1 we will 
not encounter any problems because there are no UV three-forms in our model.  
\begin{figure}[htb]\label{wilsonloop}
		\begin{center}
\includegraphics[height=6cm]{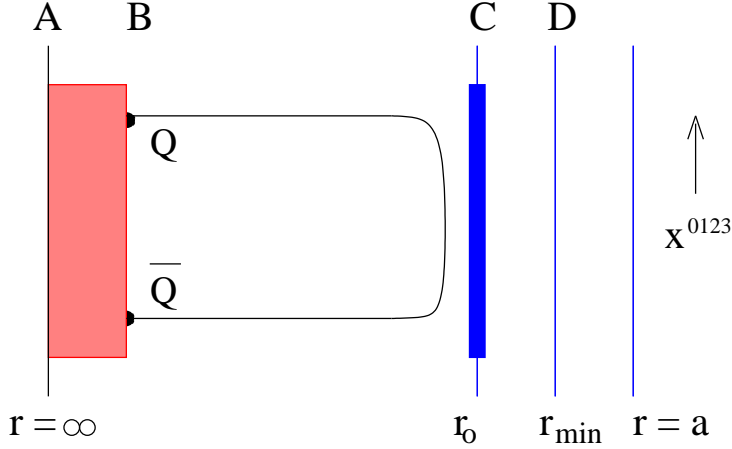}
		\caption{{The string configuration that we will use to evaluate the Wilson loop in the dual 
gauge theory. The line $A$ determines the actual boundary, with the line $B$ denoting the extent of the seven brane.
We will assume that line $B$ is very close to the line $A$. The line $C$ at $r = r_o$ denotes the boundary between 
Region 3 and Region 2. Region 2 is the interpolating region that ends at $r = r_{\rm min}$. At the far IR the geometry is
cut-off at $r = a$ from the blown-up $S^3$. As discussed in the text, the string has a maximum dip that will 
eventually lead to the confining potential between the heavy quark and the antiquark.}}
		\end{center}
		\end{figure} 

\noindent Since the system is not dynamical, the world line for the static
 $Q\bar{Q}$ can be chosen to be 
\bg \label{qline}
x^1~=~\pm \frac{d}{2},~~~~~ x^2 ~= ~ x^3 ~= ~0
\nd
and using $u \equiv 1/r$ we can rewrite the metric in Region 3 as\footnote{We will be using the Einstein summation 
convention henceforth unless mentioned otherwise.}:
\bg \label{reg3met}
ds^2&=& g_{\mu\nu} dX^\mu dX^\nu ~=~
{\cal A}_n(\psi,\theta_i,\phi_i)u^{n-2}\left[-g(u)dt^2+d\overrightarrow{x}^2\right]\nonumber\\
&+&\frac{
{\cal B}_l(\psi,\theta_i,\phi_i)u^{l}}{
{\cal A}_m(\psi,\theta_i,\phi_i)u^{m+2}g(u)}du^2+\frac{1}
{{\cal A}_n(\psi,\theta_i,\phi_i)u^{n}}~ds^2_{{\cal M}_5}
\nd
where ${\cal A}_n$ are the coefficients that can be extracted from the $a_i$ in \eqref{heaft}, the 
black hole factor $g(u) = 1$ for the zero-temperature case, 
and 
$ds^2_{{\cal M}_5}$ is the metric of the internal space that includes the corrections given in  
\eqref{rab}. This can be 
made precise as
\bg\label{cala}
{1\over \sqrt{h}}~ = ~ {1\over L^2 u^2 \sqrt{a_i u^i}} \equiv {\cal A}_n u^{n-2} ~=~ {1\over L^2 u^2}\left[a_0  
-{a_1u\over 2} + \left({3a_1^2\over 8a_0} - {a_2\over 2}\right)u^2 + ...\right]
\nd
giving us ${\cal A}_0 = {a_0\over L^2}, {\cal A}_1 = -{a_1\over 2L^2}, 
{\cal A}_2 = {1\over L^2}\left({3a_1^2\over 8a_0} - {a_2\over 2}\right)$ and so on. Note that since $a_i$, $i \ge 1$ 
are of ${\cal O}(g_sN_f)$ and $L^2 \propto \sqrt{g_sN}$, all ${\cal A}_i$ are very small.  
The $r^{-n}$ corrections along the radial direction given in \eqref{rrr} are accomodated above 
via ${\cal B}_l u^l$ series.   

Now suppose $X^\mu:(\sigma,\tau)\rightarrow (x^{0123}, u,\psi,\phi_i,\theta_i)$ is a mapping from string 
world sheet to space-time. Choosing a parametization $\tau= x^0 \equiv t,\sigma= x^1 \equiv x$ with the boundary of the 
world sheet embedding being 
the path ${\cal C}$, we see that we can have
\bg \label{ws-1}
&&X^0~= ~t,~~~ X^1~= ~x,~~~ X^2~ = ~ X^3 ~ = ~ 0, ~~~ X^7~=~u(x), ~~~X^6 ~=~\psi ~=~ 0 \nonumber\\
&&(X^4, X^5) ~ = ~ (\theta_1, \phi_1) ~ = ~ (\pi/2, 0), ~~~ 
(X^8, X^9) ~ = ~ (\theta_2, \phi_2) ~ = ~ (\pi/2, 0)
\nd
which is almost like the slice \eqref{sol} that we choose in \cite{FEP}. The advantage of such a choice is to 
get rid of the ackward angular variables that appear for our background so that we will have only a $r$ (or $u$) 
dependent background like in \eqref{slicebg} discussed before. We will also impose 
the boundary condition 
\bg \label{bc-1}
u(\pm d/2)~ = ~ u_\gamma ~\approx ~ 0
\nd
where $u_\gamma$ denote the position of the seven brane {\it closest} to the 
boundary.  The Nambu-Goto action for the string connecting this seven brane is:
\bg \label{NG-1}
&& S_{\rm string}= {T_0\over 2\pi}\int d\sigma d\tau 
\Big[\sqrt{-{\rm det}\left[(g_{\mu\nu} + \partial_\mu\phi\partial_\nu\phi)\partial_a X^\mu \partial_b X^\nu\right]} 
+ {1\over 2} \epsilon^{ab} B_{ab}
+ J(\phi)\nonumber\\
 && ~~~~~~~~+ \epsilon^{ab}
\partial_a X^m \partial_b X^n ~\bar\Theta~ \Gamma_m \Gamma^{abc....} \Gamma_n ~\Theta
~F_{abc....} + {\cal O}(\Theta^4)\Big]
\nd   
where $a, b=1,2$, $\partial_1\equiv \frac{\partial}{\partial \tau}$, $\partial_2\equiv \frac{\partial}{\partial \sigma}$. 
The other fields appearing in the action are the pull backs of the NS $B$ field $B_{ab}$, the dilaton coupling $J(\phi)$ 
and the RR field strengths
$F_{abc..}$. Its clear that once we switch off the fermions i.e $\Theta = \bar\Theta = 0$ the RR fields decouple.  
The $B_{NS}$ field do couple to the fundamental string but as we discussed before, in Region 3 we don't expect to 
see any three-form field strengths. This is because the amount of $B_{\rm NS}$ that could leak out from Region 2 to 
Region 3 is:
\bg\label{leaking} 
B_{\rm NS} ~= ~ M {\cal S} [1-f(r)] ~ = ~ M{\cal S} ~e^{-\alpha(r - r_0)}, ~~~~ r > r_0
\nd
where ${\cal S}$ is the two-form:
\bg\label{sdefin}
&&{\cal S} = g_s 
\left(b_1(r)\cot\frac{\theta_1}{2}\,d\theta_1+b_2(r)\cot\frac{\theta_2}{2}\,d\theta_2\right)\wedge e_\psi
- {g_sb_4(r)\over 12\pi}\sin\theta_2\,d\theta_2\wedge d\phi_2\\
&& +{3g_s\over 4\pi}\left[\left(1+g_sN_f -{1\over r^{2g_sN_f}} + {9a^2g_sN_f\over r^2}\right)
\log\left(\sin\frac{\theta_1}{2}\sin\frac{\theta_2}{2}\right) + b_3(r)\right]
\sin\theta_1\,d\theta_1\wedge d\phi_1 \nonumber\\
&&-{g_s\over 12\pi}\Bigg(2 -{36a^2g_sN_f\over r^2} + 9g_sN_f -{1\over r^{16g_sN_f}} - 
{1\over r^{2g_sN_f}} + {9a^2g_sN_f\over r^2}\Bigg)\sin\theta_2\,d\theta_2\wedge d\phi_2\nonumber 
\nd
and $b_n$ have been defined before. We see that not only $B_{\rm NS}$ has an inverse $r$ fall off, but also has  
a strong exponential decay because $\alpha >> 1$. This is the main reason why there are no NS or RR three-forms in 
Region 3, making our computation of the Wilson loop relatively easier compared to the pure Klebanov-Strassler case. 

On the other hand dilaton {will} couple {\it additionally}
via the $J(\phi)$ term. Although this coupling of $\phi$ is not to the 
$X^\mu$, we can still control this coupling by arranging the other seven-branes such that:
\bg\label{sevbrarr}
{\rm {\bf Re}}\left(\sum_{i=1}^{n_1} {3z^n_i\over z^n} - \sum_{j=1}^{n_2} {{\widetilde z}^n_j \over z^n}\right) 
~ < ~ \epsilon ~~~~~ {\rm for}~~ 0\le n \le m_o
\nd
with $\epsilon$ very small and $m_o$ a sufficiently big number. Under this condition the dilaton will be 
essentially constant and the axio-dilaton $\tau$  
would behave as:
\bg\label{axdbel}
\tau ~ = ~ \tau_0 + \sum_{n = 1}^\infty {{\cal C}_n \over r^n} + 
i\sum_{n > m_o}^\infty {{\cal D}_n \over r^n}
\nd 
so that its contribution to NG action can be ignored although the ${\cal B}_l u^l$ contribution still dominates, 
because the seven-branes continue to affect the geometry from their energy-momentum tensors and the axion 
charges. In this limit both string and Einstein frame metrics are identical and the background dilaton is 
\bg\label{bgdil}
\phi = {\rm log}~g_s - g_s{\cal D}_{n+m_o} u^{n+m_o} + {\cal O}(g_s^2)
\nd 
which, in the limit $g_s \to 0$, will be dominated by the 
constant term (note that $m_o$ is fixed). 
Because of this form, the NG string will see a slightly different background metric as 
evident from \eqref{NG-1}. 

Thus once the dust settles, using the metric \eqref{reg3met}
with the embedding $X^\mu$ given by \eqref{ws-1}, one can easily show that at zero temperature the NG action is 
given by:
\bg \label{NG-2} 
S_{\rm{NG}}=\frac{T}{2\pi}\int_{-{d\over 2}}^{+{d\over 2}} \frac{dx}{u^2}\sqrt{\Big({\cal A}_n u^n\Big)^2
+ \Big[{\cal B}_m u^m + 2g_s^2 {\widetilde{\cal D}}_{n+m_o} {\widetilde{\cal D}}_{l+m_o} {\cal A}_k u^{n+l+k+2m_o} 
+ {\cal O}(g_s^4)\Big]
\left(\frac{\partial u}{\partial x}\right)^2 }\nonumber\\
\nd
where we have used $\int dt=T/T_0 \equiv T$ (with $T_0 \equiv 1$ henceforth), 
${\widetilde{\cal D}}_{n+m_o} = (n+m_o){\cal D}_{n+m_o}$; 
and ${\cal A}_n, {\cal B}_n$ and ${\cal D}_{n+m_o}$
are now defined for choices of the angular coordinates
given in \eqref{ws-1}.  
The above action can be condensed if we redefine:
\bg\label{redefine}
{\cal B}_m u^m + 2g_s^2 {\widetilde{\cal D}}_{n+m_o} {\widetilde{\cal D}}_{l+m_o} {\cal A}_k u^{n+l+k+2m_o} 
+ {\cal O}(g_s^4) 
~ \equiv ~ {\cal G}_l u^l
\nd
which would mean that the constraint
equation i.e $\partial_1 T^1_1 = 0$, $T^1_1$ being the stress-tensor,
 for $u(x)$ derived from the action (\ref{NG-2}) using \eqref{redefine} can be written as 
\bg \label{EL-1}
\frac{d}{dx}\left(\frac{\left({\cal A}_n\ u^n\right)^2}{u^2\sqrt{\left({\cal A}_m\ u^m\right)^2
+ {\cal G}_m u^m\left(\frac{\partial u}{\partial x}\right)^2 }}\right) ~= ~0
\nd
implying that:
\bg\label{feqn}
\frac{\left({\cal A}_n u^n\right)^2}{u^2\sqrt{\left({\cal A}_m u^m\right)^2
+ {\cal G}_m u^m ~u'(x)^2}} ~ = ~ C_o
\nd
where $C_o$ is a constant, and 
$u'(x)\equiv\frac{\partial u}{\partial x}$. 
This constant $C_o$ can be determined in the following way:
as we have the endpoints of the string at $x=\pm d/2$, 
by symmetry the string will be U shaped and if $u_{\rm max}$ is the maximum value of $u$, 
we can define $u(0)= u_{\rm
max}$ and $u'(x=0)=0$. Plugging this in \eqref{feqn} we get:
\bg \label{C}
C_o ~= ~ \frac{{\cal A}_n u_{\rm max}^n }{u_{\rm max}^2}
\nd    
Once we have $C_o$, we can use \eqref{EL-1} to get the following simple differential equation: 
\bg \label{EL-2}
{du\over dx} ~= ~ \pm \frac{1}{C_o\sqrt{{\cal G}_m u^m}}\left[\frac{\left({\cal A}_n u^n\right)^4}{u^4}
-C_o^2\left({\cal A}_m u^m\right)^2\right]^{1/2}
\nd
which in turn can be used to write $x(u)$ as:
\bg \label{EL-3}
x(u) ~= ~ C_o \int_{u_{\rm max}}^u dw \frac{w^2\sqrt{{\cal G}_m w^m}}{\left({\cal A}_n w^n\right)^2} 
\left[1-\frac{C_o^2 w^4}{\left({\cal A}_m
w^m\right)^2}\right]^{-1/2}
\nd 
where we have used $x(u_{\rm max})=0$. Now using the boundary condition given in \eqref{bc-1} i.e $x(u=u_\gamma)=d/2$, 
and defining $w = u_{\rm max} v, \epsilon_o = {u_\gamma\over u_{\rm max}}$
we have
\bg \label{D-1}
d~ = ~2u_{\rm max} \int_{\epsilon_o}^{1} dv ~v^2\frac{\sqrt{{\cal G}_m u_{\rm max}^m v^m}
\left({\cal A}_n u_{\rm max}^n\right)}{\left({\cal A}_m u_{\rm max}^m v^m\right)^2} 
\left[1-v^4\left(\frac{{\cal A}_n u_{\rm max}^n}{{\cal A}_m u_{\rm max}^m v^m}\right)^2\right]^{-1/2}
\nd
At this stage we can assume all ${\cal A}_n > 0$. This is because
for ${\cal A}_n > 0$ we can clearly have degrees of freedom in the gauge theory growing towards UV, which is an 
expected property of models with RG flows. Of course this is done to simplify the subsequent analyses. Keeping 
${\cal A}_n$ arbitrary will also allow us to derive the linear confinement behavior, but this case will require a 
more careful analysis. We will leave this for future works. Note also that 
similar behavior is seen for the  
the Klebanov-Strassler model, and we have already discussed how degrees of freedom run in Regions 2 and 3. 
Another obvious condition is that $d$, which is the distance between the quarks, cannot be imaginary. 
From
(\ref{D-1}) we can see that the integral becomes complex for 
\bg \label{real-1}
{\cal F}(v)~\equiv~ v^4\left(\frac{{\cal A}_n u_{\rm max}^n}{{\cal A}_m u_{\rm max}^m v^m}\right)^2~ > ~ 1
\nd
whereas for ${\cal F}(v)= 1$ the integral becomes singular. Then for $d$ to be always real we must have 
\bg\label{real-2}
{\cal F}(v)~ \leq ~ 1
\nd 
We can now use, without loss of generality, ${\cal A}_0 = 1$ and ${\cal A}_1 = 0$ in units of $L^2$. 
Such a choice is of course consistent 
with supergravity solution for our background (as evident from \eqref{cala}). Therefore   
analyzing the condition (\ref{real-2}), one easily finds that we must have
\bg\label{real-3}
{1\over 2}(m+1){\cal A}_{m + 3}\ u_{\rm max}^{m + 3} ~ \leq  ~ 1
\nd
for $d$ to be real. This condition puts an upper bound on $u_{\rm max}$ and we can use this to constrain the fundamental
string to lie completely in Region 3 as depicted in Figure 5 earlier. 
Observe that for AdS spaces, ${\cal A}_n=0$ for
$n>0$ and hence there is no upper bound for $u_{\rm max}$. This is also the main reason why we see confinement 
using our background but not from the AdS backgrounds. Furthermore one might mistakenly think that generic 
Klebanov-Strassler background should show confinement because the space is physically cut-off due to the presence of a 
blown-up $S^3$. Although such a scenario implies a $u_{\rm max}$ for the fundamental string, this doesn't naturally 
lead to confinement because due to the presence of 
logarithmically varying $B_{\rm NS}$ fields there are UV divergences of the Wilson loop. These divergences {\it cannot}
be removed by simple regularization schemes \cite{cschu}.    

Coming back to \eqref{NG-2} we see that it can be further simplified.
Using (\ref{feqn}), (\ref{C}) and (\ref{EL-2}) in (\ref{NG-2}), 
we can write it as an integral over $u$:
\bg \label{NG-3}
S_{\rm NG}&=&~\frac{T}{\pi} \int_{u_\gamma}^{u_{\rm max}} {du\over u^2} \sqrt{{\cal G}_l u^l}
\left[1-\frac{C_o^2 u^4}{\left({\cal A}_m u^m\right)^2}\right]^{-1/2}\nonumber\\
&=&~\frac{T}{\pi}\frac{1}{u_{\rm max}}\int_{\epsilon_o}^1 {dv\over v^2} \sqrt{{\cal G}_m u_{\rm max}^m v^m}
\left[1-v^4\left(\frac{{\cal A}_n u_{\rm max}^n}{{\cal A}_m u^m_{\rm max}
v^m}\right)^2\right]^{-1/2} 
\nd
where in the second equality we have taken $v= u/u_{\rm max}$.

This simplified action \eqref{NG-3} is not the full story. It is also divergent in the limit $\epsilon_o \to 0$. 
We isolate the divergent part of the above integral (\ref{NG-3}) by first computing it as a function of $\epsilon_o$. The
result is
\bg
S_{\rm NG}&\equiv& S_{\rm NG}^{\rm I} + S_{\rm NG}^{\rm II} ~ = ~
\frac{T}{\pi}\frac{1}{u_{\rm max}}\int_{\epsilon_o}^1 {dv\over v^2} \sqrt{{\cal G}_m u_{\rm max}^m v^m}\nonumber\\
&+&\frac{T}{\pi}\frac{1}{u_{\rm max}}\int_{\epsilon_o}^1 {dv\over v^2} \sqrt{{\cal G}_m u_{\rm max}^m v^m}
\left\{\left[1-v^4\left(\frac{{\cal A}_n u_{\rm max}^n}{{\cal A}_m u^m_{\rm max}
v^m}\right)^2\right]^{-1/2}-1\right\}
\nd
Now by expanding $\sqrt{{\cal G}_m u_{\rm max}^m v^m}={\widetilde{\cal G}}_lv^l$ we can compute the first integral to be 
\bg \label{NG-3A}
S_{\rm NG}^{\rm I}&=&\frac{T}{\pi}\frac{1}{u_{\rm max}}\left(-{\widetilde{\cal G}}_0+\frac{{\widetilde{\cal G}}_0}
{\epsilon_o}+\sum_{l=2}\frac{{\widetilde{\cal G}}_l}{l-1}+{\cal O}(\epsilon_o)+..\right)
\nd
where ${\widetilde{\cal G}}_0 = {\cal G}_0, ~{\widetilde{\cal G}}_1 = {1\over 2} {\cal G}_1 u_{\rm max}$ 
and so on. The second integral becomes
\bg \label{NG-3B}
S_{\rm NG}^{\rm II}&=&\frac{T}{\pi}\frac{1}{u_{\rm max}}\int_0^1 {dv\over v^2} \sqrt{{\cal G}_m u_{\rm max}^m v^m}
\left\{\left[1-v^4\left(\frac{{\cal A}_n u_{\rm max}^n}{{\cal A}_m u^m_{\rm max}
v^m}\right)^2\right]^{-1/2}-1\right\}+{\cal O}(\epsilon_o^3)\nonumber\\
\nd
where the $\epsilon_o$ dependence here appears to 
${\cal O}(\epsilon^3_o)$; and we have set ${\cal G}_1=0$ without loss of generality. Now combining the
result in (\ref{NG-3A}) and (\ref{NG-3B}), we can obtain the renormalized action by subtracting the divergent term ${\cal
O}(1/\epsilon)$ in the limit $\epsilon_o \to 0$ and obtain the following result 
\bg \label{NG-4}
S_{\rm NG}^{\rm ren}&=&\frac{T}{\pi}\frac{1}{u_{\rm max}}
\Bigg\{-{\widetilde{\cal G}}_0+ \sum_{l=2}\frac{{\widetilde{\cal G}}_l}{l-1} 
- \int_0^1 {dv\over v^2} \sqrt{{\cal G}_m u_{\rm max}^m v^m} + 
{\cal O}(g_s^2) \nonumber\\
&+& \int_0^1 {dv\over v^2} \sqrt{{\cal G}_m u_{\rm max}^m v^m}
\left[1-v^4\left(\frac{{\cal A}_n u_{\rm max}^n}{{\cal A}_m u^m_{\rm max}
v^m}\right)^2\right]^{-1/2} + {\cal O}(\epsilon_o)\Bigg\}
\nd
where the third term in \eqref{NG-4}, including the ${\cal O}(g_s^2)$ correction,  
is related to the action for a straight string in this background in the limit 
$g_s \to 0$. Our
subtraction scheme is more involved because the straight string sees a complicated metric due to the background 
dilaton and non-Ricci flat unwarped metric. This effect is {\it independent} of any choice of the warp factor. We 
expect this action to be finite in the limit $\epsilon_o \to 0$. 

Once we have the action,
we should use this to compute the $Q\bar{Q}$ potential through (\ref{Vqq}). 
Looking at (\ref{D-1}) we observe that the relation between $d$ and $u_{\rm max}$ is parametric and can be 
quite involved depending on the
coefficients ${\cal A}_n$. 
If we have ${\cal A}_n=0, {\cal G}_n=0$ for $n>0$, 
we recover the well known AdS result, namely: $d\sim u_{\rm max}$ and $V_{Q\bar{Q}}\sim {1\over d}$. But in
general (\ref{D-1}) and (\ref{NG-4}) should be solved together to obtain the potential.  

As it stands, \eqref{D-1} and \eqref{NG-4} are both rather involved. So to find some correlation between them we
need to go to the limiting behavior of $u_{\rm max}$. Therefore in the following,  
we will study the behaviour of $d$
and $S_{\rm NG}^{\rm ren}$ for the cases where $u_{\rm max}$ is large and small.

\subsubsection{Quark-Antiquark potential for small $u_{\rm max}$}

Let us first consider the case where $u_{\rm max}$ is small.  In this limit we can impose 
$u_\gamma = \epsilon u_{\rm max}$, so that $\epsilon_o = \epsilon$ in all the above integrals and 
consequently their lower limits 
will be independent of $u_{\rm max}$. 
We can also approximate
\bg\label{lbaz}
{\cal A}_n u^n_{\rm max} ~=~ {\cal A}_0 ~+~ {\cal A}_2 u^2_{\rm max} ~ \equiv~ 1 ~+~ \eta
\nd
where ${\cal A}_0 = 1$ and ${\cal A}_2 u^2_{\rm max}=\eta$. Using this we can 
 write both (\ref{D-1}) and (\ref{NG-4}) as Taylor series in $\eta$ around $\eta=0$. The result is
\bg \label{D-3}
&& d ~ = ~ \sqrt{\eta}\left[a_0 ~ + ~ a_1\eta + {\cal O}(\eta^2)\right]\nonumber\\
&& S_{NG}^{\rm ren} ~ = ~ {T\over \pi} \left[{b_0 + b_1\eta + {\cal O}(\eta^2)\over \sqrt{\eta}}\right]
\nd 
with $a_0, a_1, b_0, b_1$ are defined in the following way:
\bg\label{abdefn}
&&a_0 ~=~ {2\over \sqrt{{\cal A}_2}} \int_0^1 dv {v^2\over \sqrt{1-v^4}} ~=~ {1.1981\over \sqrt{{\cal A}_2}}\nonumber\\
&& a_1 ~=~ 
{2\over \sqrt{{\cal A}_2}} \int_0^1 dv {v^2\over \sqrt{1-v^4}}\left[{1-v^6\over 1-v^4} + \left({{\cal G}_2 - 
4{\cal A}_2\over 2{\cal A}_2}\right)v^2\right]\nonumber\\
&&b_0 ~ = ~ \sqrt{{\cal A}_2}\left[-1 +
\int_0^1 dv \left({1-\sqrt{1-v^4} \over v^2\sqrt{1-v^4}}\right)\right] ~ = ~ -0.62 \sqrt{{\cal A}_2}\nonumber\\
&& b_1 ~ = ~ {1\over 2\sqrt{{\cal A}_2}}\Bigg\{{\cal G}_2 +
\int_0^1 dv \left[{2{{\cal A}_2} v^4 + {\cal G}_2 v^2(1+v^2)(1-
\sqrt{1-v^4})\over v^2(1+v^2)\sqrt{1-v^4}}\right]\Bigg\}
\nd
where we have taken ${\cal G}_0 = 1$ and ${\cal G}_1 = 0$ without loss of generality. Note that all the above 
integrals are independent of $\eta$ (or $u_{\rm max}$) because all 
${\cal O}(\epsilon)$ corrections are independent of $u_{\rm max}$. Note also that $b_0 = -\vert b_0\vert$.  
In this limit clearly increasing $\eta$ increases $d$, the distance between the quarks. For small $\eta$, 
$d = a_0\sqrt{\eta}$, and therefore the Nambu-Goto action will become:
\bg\label{NGac}
S_{\rm NG}^{\rm ren}  ~ = ~ T\left[-\left({a_0 \vert b_0\vert\over \pi}\right) {1\over d} 
~+~ \left({b_1\over \pi a_0}\right)d
+ {\cal O}(d^3)\right]
\nd
where all the constants have been defined in \eqref{abdefn}. Using \eqref{Vqq} we can determine the short-distance 
potential to be (recall $T_0 = 1$):
\bg\label{sdpot}
V_{Q\bar{Q}} ~&& = ~ -\left({a_0\vert b_0\vert\over \pi}\right) {1\over d} ~+~ \left({b_1\over \pi a_0}\right)d
+ {\cal O}(d^3)\nonumber\\
&&= ~ -{0.236\over d} ~+~\left(0.174{\cal G}_2 + 0.095{\cal A}_2\right) d ~ + ~ {\cal O}(d^3)
\nd
which is dominated by the inverse $d$ behavior, i.e the expected Coulombic behavior. Note that the coefficient of the 
Coulomb term is independent of the warp factor and therefore should be universal. This result, in appropriate units, 
is of the same order of magnitude as the real Coulombic term for the Charmonium 
spectra \cite{karsch, charmonium, brambilla, boschi}. This 
prediction, along with the overall minus sign, should be regarded as a success of our model (see also \cite{zakahrov}
where somewhat similar results have been derived in a string theory inspired model). The second term on the other 
hand is model dependent, and vanishes in the pure AdS background. 

Note also that the above computations are valid for infinitely massive quark-antiquark pair. For lighter quarks, we 
expect the results to differ. It would be interesting to compare these results with the ones where quarks are 
much lighter.  
 
\subsubsection{Quark-Antiquark potential for large $u_{\rm max}$}

To analyse the quark-antiquark potential for large $u_{\rm max}$ we first define a quantity called $z_{\rm max} 
\equiv {\widetilde u}^{-1}_{\rm max}$ which would be our small tunable parameter. We note that the {\it smallest} 
$z_{\rm max}$ will come from the following equality: 
\bg\label{real-8}
{1\over 2}\sum_{m}(m+1){{\cal A}_{m + 3}\over z_{\rm max}^{m + 3}} ~ = ~ 1
\nd 
which is the upper bound on the inequality \eqref{real-3}. Furthermore since we demanded ${\cal A}_n \ge 0$,
the above 
condition will imply that the coefficients ${\cal A}_n$ has to quickly become very small because in the 
limit $z_{\rm max} < 1$
\bg\label{condp}
\lim_{m\to \infty} ~{m+1\over z_{\rm max}^{m + 3}} ~\to ~\infty
\nd
which is perfectly consistent with \eqref{cala} because higher ${\cal A}_n$ are proportional to higher 
powers of $g_sN_f$ and therefore strongly suppressed in the limit $g_sN_f \to 0$.
This will also mean that we can retain only few of the ${\cal A}_n$'s to study the potential for small $z_{\rm max}$.
In fact we will soon give an estimate of the largest $n$ that we should keep in our analysis. 

To determine the distance between the quark we can again use \eqref{EL-3}. However we have to be careful about a few
subtleties that appear due to our choice of the scale $z_{\rm max}$. First of all note that we will use 
$w = z_{\rm max} v$ in \eqref{EL-3}. This will immediately imply that the lower bound of the integral is no longer
$\epsilon$ that we had in the previous subsection, but it is 
\bg\label{lbe}
{u_\gamma\over z_{\rm max}} ~ = ~ {\epsilon u_{\rm max} \over z_{\rm max}} ~ = ~ {\widetilde\epsilon} ~ \to ~0
\nd
where $u_{\rm max}$ is the lowest value from the inequality \eqref{EL-3} that we chose in the previous subsection (to 
avoid clutter we use the same notation). Note that ${\widetilde\epsilon}$ is independent of $z_{\rm max}$. Using
this we can now write the distance $d$ between the quark and the antiquark as:
\bg\label{dbqaq}
d ~ = ~ 2V_0 z^5_{\rm max} \int_{\widetilde\epsilon}^{1/z^{2}_{\rm max}} dv ~v^2
~{\sqrt{{\cal G}_m v^m z^m_{\rm max}} \over
({\cal A}_nv^nz^n_{\rm max})^2} \Bigg[1 - z^8_{\rm max}{V_0^2\over  ({\cal A}_kv^kz^k_{\rm max})^2}\Bigg]^{-1/2}
\nd
where we have defined $V_0 = {\cal A}_n z^{-n}_{\rm max}$ and sum over repeated index is implied as before.

Comparing $V_0$ with \eqref{real-8} we see that $V_0$ can be made small if ${\cal A}_2 << 1$ (which is consistent with 
\eqref{cala} and \eqref{lbaz}) as all ${\cal A}_n$ for $n \ge 3$ are very small. 
Additionally, from 
\eqref{dbqaq}, the term $z^8_{\rm max} V_0^2$ will imply that it will be sufficient to restrict $V_0$ to the following 
series:
\bg\label{v0zseries}
V_0 ~ = ~ 1 ~+~ {{\cal A}_2 \over z^2_{\rm max}}~+~ {{\cal A}_3 \over z^3_{\rm max}}~+~ {{\cal A}_4 \over z^4_{\rm max}}
\nd
because ${\cal A}_5$ onwards are very small to consistently
maintain the reality of $d$ in \eqref{dbqaq} as well as their $g_sN_f$ dependences from \eqref{cala}. This means $d$ 
in \eqref{dbqaq} can be further simplified to:
\bg\label{dbeq}
d  &&= ~ 2V_0 z^5_{\rm max} \int_{\widetilde\epsilon}^{1/z^{2}_{\rm max}} dv ~v^2 ~~
{1-\left({\cal A}_2 -{1\over 2}{\cal G}_2\right) z_{\rm max}^2 v^2 \over \sqrt{1 - z_{\rm max}^8V_0^2 + 
2 z_{\rm max}^{10} V_0^2 {\cal A}_2 v^2}}\\
&& \approx ~ 2V_0\Bigg\{\left(1+ {1\over 2}z_{\rm max}^8 V_0^2\right){1\over 3 z_{\rm max}} + \left[{1\over 2}
\left({\cal G}_2 - 4{\cal A}_2\right)z_{\rm max}^2 + {1\over 4}\left({\cal G}_2 - 8{\cal A}_2\right)V_0^2
z_{\rm max}^{10}\right]{1\over 5 z_{\rm max}^5}\Bigg\}\nonumber
\nd 
Since we have taken both ${\cal A}_2$ as well as ${\cal B}_2$ to be very small, and plugging in the value of $V_0$ 
from \eqref{v0zseries} we see that $d$ is dominated mostly by inverse $z_{\rm max}$ terms, i.e
\bg\label{ddom}
d ~ = ~ {2\over 3z_{\rm max}}\left[1+{{\cal A}_4^2\over 2} + {\cal O}({\cal A}_n^3)\right] + 
{2\over 3z_{\rm max}^2} {\cal O}({\cal A}_n^2) + {\cal O}\left({1\over z_{\rm max}^3}\right)
\nd 
The renormalised Nambu-Goto action on the other hand takes the following form:
\bg\label{rnnmo}
S_{\rm NG}^{\rm ren} && =~ {T\over \pi z_{\rm max}}\int_0^{1/z_{\rm max}^2}~{dv\over v^2} 
\sqrt{{\cal G}_l z^l_{\rm max} v^l}\Bigg[
\left(1 - z^8_{\rm max}{V_0^2\over  ({\cal A}_kv^kz^k_{\rm max})^2}\right)^{-1/2} -1\Bigg]\nonumber\\
&& ~~~ + {T\over \pi z_{\rm max}}\left[-z_{\rm max}^2 + {{\cal G}_2\over 2} + 
{{\cal G}_3\over 4}{1\over z_{\rm max}} + {8{\cal G}_4 - {\cal G}_2^2\over 48} {1\over z_{\rm max}^2} + ...
\right]\\
&& \approx ~ -{T\over 2\pi} (2 + {\cal A}_4^2) z_{\rm max} 
- {T{\cal A}_4^2\over 2\pi}\left[2{\cal A}_2 - {{\cal G}_2}\left({{\cal A}_3^2\over {\cal A}_4^2} + 
2{{\cal A}_2\over {\cal A}_4}\right) 
 - {{\cal G}_2\over {\cal A}_4^2}
\right] ~{1\over z_{\rm max}} ~+ ~ {\cal O}\left({1\over z_{\rm max}^2}\right)\nonumber
\nd
where its clear that the string action is dominated by the inverse $z_{\rm max}$ terms. Now substituting \eqref{ddom} 
and \eqref{rnnmo} in \eqref{Vqq}, we get 
\bg\label{linpot}
V_{Q\bar Q} ~ = ~  {3{\cal A}^2_4\over 4\pi}\left[2{\cal A}_2 - {{\cal G}_2}\left({{\cal A}_3^2\over {\cal A}_4^2} + 
2{{\cal A}_2\over {\cal A}_4}\right) 
 - {{\cal G}_2\over {\cal A}_4^2} + ... \right]~d + {\cal O}\left({1\over d}\right) 
\nd
which is the required linear potential between the quark and the antiquark. The above potential can also be 
rewritten as:
\bg\label{potred}
V_{Q\bar Q} ~ = ~ \left({{\cal H}_n \alpha^n_{\rm max}\over \pi \alpha^2_{\rm max}}\right) d + 
{\cal O}\left({1\over d}\right)
\nd
where $\alpha_{\rm max} \equiv {1\over {\cal A}_4}$ and 
${\cal H}_0 = {3{\cal A}_2\over 2}, 
{\cal H}_1 = -{3{\cal A}_2 {\cal G}_2\over 2}, {\cal H}_2 = -{3{\cal G}_2\over 4}(1+{\cal A}_3^2)$ etc. It will soon
become clearer why we want to express the potential \eqref{potred} in this way.

However there is one nagging issue that might be bothering the reader, namely, how do we know that the potential
\eqref{potred} or equivalently \eqref{linpot} only has a linear term? To answer this question convincingly, we go to the 
next section where we provide a generic derivation of the linear term.

\subsubsection{Generic argument for confinement}

In the above subsection we argued for the linear potential taking all ${\cal A}_n$ for $n \ge 1$ to be small. This is 
consistent with the {\it supergravity} limit of our background because in this limit we expect $g_s \to 0$ and 
$g_sN_f \to 0$ with $g_sN \to \infty$. For these choices of ${\cal A}_n$, \eqref{real-8} will imply $z_{\rm max} < 1$ 
because we expect ${\widetilde u}_{\rm max}$ to take the highest value in Region 3. Under such a situation, 
condition like \eqref{condp} will be fully under control, and an analysis of the Wilson loop 
above reproduces the required 
linear potential at large $d$. 

However a little thought will tell us that the above derivation cannot be the complete story. What if $u_{\rm max}$, in
appropriate units,  
is of order ($1-\epsilon$) where $\epsilon \to 0$? 
In that case its inverse $z_{\rm max}$ is of order 1, so both 
$u_{\rm max}$ and $z_{\rm max}$ can no longer be good expansion parameters. 
We may also consider simultaneously the case where 
$g_s$ is no longer small so that ${\cal A}_n$ for $n \ge 1$ are not small either. Such choices will take us 
away from the supergravity limit that we have been following. In this limit, we want to ask whether we can still 
show linear confinement of quarks. Or more generically we want to study confinement for a choice of $u_{\rm max}$ that 
saturates the inequality \eqref{real-3} but does not presuppose any limiting behavior of  $u_{\rm max}, {\cal A}_n$ or 
${\cal G}_n$. 

In the following therefore 
we will analyze the integrals (\ref{D-1}) and (\ref{NG-4}) in the limit $u_{\rm max}$ is close to it's upper bound set
by (\ref{real-3}) (see also \cite{zakahrov}). In particular if ${\bf u}_{\rm max}$ 
is the upperbound of $u_{\rm max}$, then it is found by solving
\bg \label{real-4}
{1\over 2}(m+1){\cal A}_{m + 3} {\bf u}_{\rm max}^{m + 3} ~ = ~ 1
\nd 
We observe that both the integrals (\ref{D-1}) and (\ref{NG-4}) are
dominated by $v\sim 1$ behaviour of the integrands. Near $v=1$ and 
$u_{\rm max}\rightarrow {\bf u}_{\rm max}$ the distance $d$ between the quark and the antiquark can be written as:
\bg \label{D-4}
d&& =~~2\frac{\sqrt{{\cal G}_m {\bf u}_{\rm max}^m}{\bf u}_{\rm max}}{{\cal A}_n {\bf u}_{\rm max}^n} 
\int_0^{1} \frac{dv}{\sqrt{{\bf A} (1-v)+{\bf
B}(1-v)^2}}\nonumber\\
&& = -2 \frac{\sqrt{{\cal G}_m {\bf u}_{\rm max}^m}{\bf u}_{\rm max}}{{\cal A}_n {\bf u}_{\rm max}^n} 
\left[{{\rm log}{\bf A}-{\rm log}\left(2\sqrt{{\bf B}({\bf A}+{\bf B})}+2{\bf B}+{\bf A}\right)\over \sqrt{\bf B}}\right]
\nd
where note that we have taken the lower limit to 0. This will not change any of our conclusion as we would soon see. On  
the other hand, the renormalised Nambu-Goto action for the string now becomes:
\bg \label{NG-6}
S_{\rm NG}^{\rm ren}&&= ~ {T\over \pi}\frac{\sqrt{{\cal G}_m {\bf u}_{\rm max}^m}}{{\bf u}_{\rm max}}
\left[\int_0^{1} \frac{dv}{\sqrt{{\bf A} (1-v)+{\bf
B}(1-v)^2}} -1\right] - {T\over \pi {\bf u}_{\rm max}} + {\cal O}({\bf u}^2_{\rm max})\nonumber\\
&& = ~ -{T\over \pi}\frac{\sqrt{{\cal G}_m {\bf u}_{\rm max}^m}}{{\bf u}_{\rm max}}
\left[{{\rm log}{\bf A}-{\rm log}\left(2\sqrt{{\bf B}({\bf A}+{\bf B})}+2{\bf B}+{\bf A}\right)\over \sqrt{\bf B}} 
-1\right]\nonumber\\
&& ~~~~~~ - {T\over \pi {\bf u}_{\rm max}} + {\cal O}({\bf u}^2_{\rm max})
\nd
where ${\bf A}$ and ${\bf B}$ are defined as:
\bg \label{AB}
{\bf A}&&= ~ 4 - 2\frac{n{\cal A}_n {u}_{\rm max}^n}{{\cal A}_m {u}_{\rm max}^m}\\
{\bf B}&& = ~ 8\frac{n{\cal A}_n {u}_{\rm max}^n}{{\cal A}_m {u}_{\rm max}^m}
-3\left(\frac{n{\cal A}_n {u}_{\rm max}^n}{{\cal A}_m {u}_{\rm max}^m}\right)^2
+ \frac{(n^2-n){\cal A}_n {u}_{\rm max}^n}{{\cal A}_m{u}_{\rm max}^m} - 6\nonumber
\nd
Observe that in the integral (\ref{D-4}) and (\ref{NG-6}) we have to take the limit 
$u_{\rm max}\rightarrow {\bf u}_{\rm max}$. 
So ${\bf A},{\bf B}$ should be evaluated in the same limit. Interestingly, comparing \eqref{AB} to \eqref{real-4}
we see that
\bg \label{AB1}
\lim_{u_{\rm max}\rightarrow {\bf u}_{\rm max}}~{\bf A}\rightarrow 0 
\nd
thus vanishes when computed exactly at ${\bf u}_{\rm max}$. The other quantity ${\bf B}$ remains finite at that point and
in fact behaves as:
\bg\label{bbeh}
{\bf B} = {n^2{\cal A}_n {\bf u}^n_{\rm max} \over {\cal A}_m {\bf u}^m_{\rm max}} - 4 ~ > ~0
\nd
Our above computation would  mean that the distance $d$ between the quark and the antiquark, and the Nambu-Goto 
action will have the following dominant behavior:
\bg\label{dNG}
d && = ~ \lim_{\epsilon\to 0}~
\frac{2\sqrt{{\cal G}_m {\bf u}_{\rm max}^m}{\bf u}_{\rm max}}{{\cal A}_n {\bf u}_{\rm max}^n} 
~ {{\rm log}~\epsilon \over \sqrt{\bf B}}\nonumber\\
S_{\rm NG}^{\rm ren}&& = ~ \lim_{\epsilon\to 0}~ 
{T\over \pi}\frac{\sqrt{{\cal G}_m {\bf u}_{\rm max}^m}}{{\bf u}_{\rm max}}~
 {{\rm log}~\epsilon \over \sqrt{\bf B}}
\nd
which means both of them have identical logarithmic divergences. Thus the {finite} quantity is the {\it ratio} between
the two terms in \eqref{dNG}. This gives us:
\bg\label{ratio} 
{S_{\rm NG}^{\rm ren}\over d} ~ = ~ {T\over \pi}{{\cal A}_n {\bf u}_{\rm max}^n \over {\bf u}_{\rm max}^2} ~ = ~ 
T \times {\rm constant}
\nd
Now using the identity \eqref{Vqq} and the above relation \eqref{ratio} we get our final result:
\bg\label{lipo}
V_{Q\bar Q} ~ = ~ \left({{\cal A}_n {\bf u}_{\rm max}^n \over \pi {\bf u}_{\rm max}^2}\right) ~d
\nd
which is the required linear potential between the quark and the antiquark. Observe that the above potential
has exactly the same form as \eqref{potred}, justifying the fact that ${\cal O}(d^2)$ terms are not generated 
for this case. 

Before we end this section one comment is in order. The result for linear confinement only depends on the existence
of ${\bf u}_{\rm max}$ which comes from the constraint equation \eqref{real-4}. We have constructed the background 
such that ${\bf u}_{\rm max}$ lies in Region 3, although a more generic case is essentially doable albeit technically 
challenging without necessarily revealing new physics. 
For example when ${\bf u}^{-1}_{\rm max}$ is equal to the size of the blown up $S^3$ at the IR will require
us to consider a Wilson loop that goes all the way to Region 1. The analysis remains somewhat identical to what we did 
before except that in Regions 2 and 1 we have to additionally consider $B_{\rm NS}$ fields of the form 
$u^{\epsilon_{(\alpha)}}$ and ${\rm log}~u$ respectively. Of course both the metric and the dilaton will also have 
non-trivial $u$-dpendences in these regions. One good thing however is that the Wilson loop computation have 
no UV or IR divergences whatsoever despite the fact that now the 
analysis is technically more challenging. Our expectation would be to get similar linear behavior as \eqref{lipo} 
here too. 
We will however leave a more detailed exposition of this for future works.

\subsection{Computing the Nambu-Goto Action: Non-Zero Temperature}

After studying the zero temperture behavior it is now time to discuss the case when we switch on a non-zero 
temperature i.e make $g(u) < 1$ or equivalently the inverse horizon radius, $u_h$ finite in \eqref{reg3met}, where 
\bg\label{g}
g(u)=1-{u^4\over u_h^4}
\nd 
Choosing the same quark world line (\ref{qline}) and the string embedding (\ref{ws-1}) with the same boundary condition
(\ref{bc-1}) but now in Euclidean space with compact time direction, the string action at 
finite temperature can be written as
\bg \label{NGfinT} 
S_{\rm{NG}}=\frac{T}{2\pi}\int_{-{d\over 2}}^{+{d\over 2}} \frac{dx}{u^2}\sqrt{g(u)\Big({\cal A}_n u^n\Big)^2
+ \Bigg[{\cal G}_m u^m - {2g_s^2 {\widetilde{\cal D}}_{n+m_o} {\widetilde{\cal D}}_{l+m_o} 
{\cal A}_k u^{4+n+l+k+2m_o} \over u_h^4}\Bigg]
\left(\frac{\partial u}{\partial x}\right)^2 }\nonumber\\
\nd
where ${\cal G}_m u^m$ is defined in \eqref{redefine} and the correction to ${\cal G}_m u^m$ is suppressed by 
$g_s^2$ as well as $u^4/u_h^4$ because the background dilaton and non-zero temperature induces a slightly 
different world-sheet metric than what one would have naively taken. To avoid clutter, one can further redefine
these corrections as:
\bg\label{redefagain}
{\cal G}_m u^m - {2g_s^2 {\widetilde{\cal D}}_{n+m_o} {\widetilde{\cal D}}_{l+m_o} 
{\cal A}_k u^{4+n+l+k+2m_o} \over u_h^4} ~\equiv~ {\widetilde{\cal D}}_l u^l
\nd
Minimizing this action gives the equation of motion for $u(x)$ and using the exact same procedure 
as for zero temperature,
the corresponding equation for the distance between the quarks can be written as:
\bg \label{DT-1}
d&=&2u_{\rm max} \int_{0}^{1} dv \Bigg\{v^2 \sqrt{{\widetilde{\cal D}}_m u_{\rm max}^m v^m}
\frac{\sqrt{1-\frac{u_{\rm max}^4}{u_h^4}}{\cal A}_n u_{\rm max}^n}
{\left(1-\frac{v^4u_{\rm max}^4}{u_h^4}\right)\left({\cal A}_m u_{\rm max}^m v^m\right)^2}\nonumber\\
&& 
\left[1-v^4\frac{\left(1-\frac{u_{\rm max}^4}{u_h^4}\right)}{\left(1-\frac{v^4u_{\rm max}^4}{u_h^4}\right)}
\left(\frac{{\cal A}_n u_{\rm max}^n}{{\cal A}_m u_{\rm max}^m v^m}\right)^2\right]^{-1/2}\Bigg\}
\nd
Once we have $d$, the 
renormalized Nambu-Goto action can also be written following similar procedure. The result is
\bg \label{NGT-3}
S_{\rm NG}^{\rm ren}&=&\frac{T}{\pi}\frac{1}{u_{\rm max}}
\Bigg\{-{\widehat{\cal D}}_0+ \sum_{l=2}\frac{{\widehat{\cal D}}_l}{l-1} 
- \int_0^1 {dv\over v^2} \sqrt{{\widetilde{\cal D}}_m u_{\rm max}^m v^m} + 
{\cal O}(g_s^2)\\
&+& \int_0^1 {dv\over v^2} \sqrt{{\widetilde{\cal D}}_m u_{\rm max}^m v^m}
\left[1-v^4\frac{\left(1-\frac{u_{\rm max}^4}{u_h^4}\right)}{\left(1-\frac{v^4u_{\rm max}^4}{u_h^4}\right)}
\left(\frac{{\cal A}_n u_{\rm max}^n}{{\cal A}_m u^m_{\rm max}
v^m}\right)^2\right]^{-1/2} + {\cal O}(\epsilon_o)\Bigg\}\nonumber
\nd
which is somewhat similar in form with \eqref{NG-4}, which we reproduce in the limit $u_h \to \infty$. 
Also as in \eqref{NG-4}, we have defined 
$\sqrt{{\widetilde{\cal D}}_m u_{\rm max}^m v^m} \equiv {\widehat{\cal D}}_l v^l$. 

Now just like the zero temperature case, requiring that $d$ be real, 
sets an upper bound to $u_{\rm max}$, that we 
denote again by ${\bf u}_{\rm max}$, and is found by solving the following equation:
\bg\label{real-5}
{1\over 2} (m+ 1){\cal A}_{m+3} {\bf u}_{\rm max}^{m+3}
+ {1\over j!}\prod_{k=0}^{j-1}\left(k-\frac{1}{2}\right)\left(\frac{{\bf u}_{\rm max}^4}{u_h^4}\right)^j
\left[{\cal A}_{l}{\bf u}_{\rm max}^{l}\left({l\over 2}+ 2j - 1\right)\right] ~= ~1\nonumber\\
\nd
Once we fix $u_h$ and the coefficients of the warp factor ${\cal A}_n$, ${\bf u}_{\rm max}$ will be known. We will 
assume that ${\bf u}_{\rm max}$ lies in Region 3. 

Rest of the analysis is almost similar to the zero temperature case, although the final conclusions would be quite 
different. To proceed further, let us define certain new variables in the following way:
\bg\label{newvar}
{\widetilde{\cal A}}_l ~& = & ~ \sum_m {{\cal A}_m\over u_h^{l-m}} {1\over \left({l-m\over 4}\right)!} 
\prod_{k=0}^{{l-m\over 4}-1} \left(k - {1\over 2}\right), ~~~~~ l-m \ge 4 \nonumber\\
&= & ~ 0 ~~~~~~~~~~~~~~~~~~~~~~~~~~~~~~~~~~~~~~~~~~~~~~~ l-m < 4\nonumber\\
& = & ~ {\cal A}_l ~~~~~~~~~~~~~~~~~~~~~~~~~~~~~~~~~~~~~~~~~~~~~ l - m = 0
\nd
As before, we observe that for $u_{\rm max}\rightarrow {\bf u}_{\rm max}$, 
both the integrals (\ref{DT-1}),(\ref{NGT-3}) are dominated by the behaviour of the integrand near $v\sim 1$, 
where we can write 
\bg\label{dT}
d&& =~~2\frac{\sqrt{{\widetilde{\cal D}}_m {\bf u}_{\rm max}^m}{\bf u}_{\rm max}}
{\sqrt{1-{{\bf u}_{\rm max}^4\over {u}^4_{h}}}{\cal A}_n {\bf u}_{\rm max}^n} 
\int_0^{1} \frac{dv}{\sqrt{{\widetilde{\bf A}} (1-v)+ {\widetilde{\bf
B}}(1-v)^2}}\nonumber\\ 
&& = -2 \frac{\sqrt{{\widetilde{\cal D}}_m {\bf u}_{\rm max}^m}{\bf u}_{\rm max}}
{\sqrt{1-{{\bf u}_{\rm max}^4\over {u}^4_{h}}}{\cal A}_n {\bf u}_{\rm max}^n} 
\left[{{\rm log}{\widetilde{\bf A}}-{\rm log}\left(2\sqrt{{\widetilde{\bf B}}({\widetilde{\bf A}}+{\widetilde{\bf B}})}
+2{\widetilde{\bf B}}+{\widetilde{\bf A}}\right)\over \sqrt{\widetilde{\bf B}}}\right]
\nd
where taking the lower limit of the integral to 0 again do not change any of our conclusion. On  
the other hand, the renormalised Nambu-Goto action for the string now becomes:
\bg \label{NGT}
S_{\rm NG}^{\rm ren}&&= ~ {T\over \pi}\frac{\sqrt{{\widetilde{\cal D}}_m {\bf u}_{\rm max}^m}}{{\bf u}_{\rm max}}
\left[\int_0^{1} \frac{dv}{\sqrt{{\widetilde{\bf A}} (1-v)+ {\widetilde{\bf
B}}(1-v)^2}} -1\right] - {T\over \pi {\bf u}_{\rm max}} + {\cal O}({\bf u}^2_{\rm max})\nonumber\\
&& = ~ -{T\over \pi}\frac{\sqrt{{\widetilde{\cal D}}_m {\bf u}_{\rm max}^m}}{{\bf u}_{\rm max}}
\left[{{\rm log}{\widetilde{\bf A}}-{\rm log}\left(2\sqrt{{\widetilde{\bf B}}({\widetilde{\bf A}}+{\widetilde{\bf B}})}
+2{\widetilde{\bf B}}+{\widetilde{\bf A}}\right)\over \sqrt{\widetilde{\bf B}}} 
-1\right]\nonumber\\
&& ~~~~~~ - {T\over \pi {\bf u}_{\rm max}} + {\cal O}({\bf u}^2_{\rm max})
\nd
where ${\widetilde{\bf A}}$ and ${\widetilde{\bf B}}$ are defined exactly as in \eqref{AB} but with ${\cal A}_n$ replaced
by ${\widetilde{\cal A}}_n$ given by \eqref{newvar} above. It is also clear that:
\bg \label{ABnm}
\lim_{u_{\rm max}\rightarrow {\bf u}_{\rm max}}~{\widetilde{\bf A}}\rightarrow 0 
\nd
and so both \eqref{dT} as well as \eqref{NGT} have identical logarithmic divergences. This would imply that the 
finite quantity is the ratio between \eqref{NGT} and \eqref{dT}:
\bg\label{rationow} 
{S_{\rm NG}^{\rm ren}\over d} ~ = ~ {T\over \pi}
\left({1-{{\bf u}_{\rm max}^4\over {u}^4_{h}}}\right)^{1\over 2}{{\cal A}_n {\bf u}_{\rm max}^n 
\over {\bf u}_{\rm max}^2}
\nd 
Now using the identity \eqref{wlfe} and the above relation \eqref{ratio} we get our final result:
\bg\label{liponow}
V_{Q\bar Q} ~ = ~ {\sqrt{1-{{\bf u}_{\rm max}^4\over {u}^4_{h}}}}\left({{\cal A}_n {\bf u}_{\rm max}^n 
\over \pi {\bf u}_{\rm max}^2}\right) ~d
\nd 

\subsubsection{Analysis of the melting temperature}

To determine the behavior of the potential $V_{Q\bar Q}$ as the temperature is raised or decreased (or $u_h$ is 
decreased or increased respectively) we have to carefully analyse the behavior of ${\bf u}_{\rm max}$ as a function of
$u_h$. 
Comparing this with \eqref{real-5} and fixed ${\cal A}_n$ we observe that 
\bg\label{diffva}
{\delta{\bf u}_{\rm max} \over \delta u_h} ~=~ {{1\over j!} \prod \left(k - {1\over 2}\right) 
{{\bf u}_{\rm max}^{4j}\over 
u_h^{4j+1}} {\cal A}_l u^l \left({l\over 2} + 2j -1\right)\over {1\over 2} m(m+1) {\cal A}_m u^m + 
{1\over j!} \prod \left(k - {1\over 2}\right) {{\bf u}_{\rm max}^{4j-1}\over u_h^{4j}} {\cal A}_l u^l 
(4j+l)\left({l\over 2} + 2j -1\right)}
\nd
where the repeated indices are all summed over and the product runs from $k = 0$ to $k = j-1$. Observe that the numerator
of \eqref{diffva} is always negative, and for large $u_h$ the denominator will be positive (because we are taking 
all ${\cal A}_n > 0$). This means $${\delta{\bf u}_{\rm max} \over \delta u_h}~ < ~ 0$$ and therefore
as $u_h$ is decreased (i.e the temperature is increased), ${\bf u}_{\rm max}$ increases making the 
ratio ${u_{\rm max}^4\over u_h^4}$ to increase. This in turn would imply that the slope of the potential $V_{Q\bar Q}$
decreases. Therefore there would be a temperature where the slope would be minimum and the system would 
show the property of {\it melting}. 

To start off let us consider \eqref{real-5} for a simple case where we keep $u_{\rm max}$ only to quartic 
order\footnote{This is a subtle issue because we are {\it truncating} the series \eqref{real-5} 
and especially $u_h$ to only this order to have an analytic control on our calculations. A more generic analysis 
can be done numerically, which we present in the next section.}. 
This 
means:
\bg\label{quartico}
{1\over 2} {\cal A}_3 u_{\rm max}^3 ~+~ \left({\cal A}_4 - {1\over 2u_h^4}\right)u_{\rm max}^4 ~ = ~ 1
\nd
This is a quartic equation and one can easily solve it for $u_{\rm max}$. To make the analysis a little more 
simpler, let us also assume ${\cal A}_3 = 0$. Such a choice will immediately give us the following potential 
between the quark and the antiquark:
\bg\label{simpol}
V_{Q\bar Q} ~ = ~ \left[\left(1+ {{\cal A}_2\over \pi}\right) + {1\over \pi} \sqrt{{\cal A}_4 - {1\over 2u_h^4}}\right]d
\nd
which tells us that $u_h$ has to be bounded by $$u_h ~ > ~ {0.84\over {\cal A}^{1/4}_4}$$ for 
\eqref{simpol} to make sense, and the 
slope of the potential would decrease as $u_h$ approaches this value. On the other hand $u_{\rm max}$ increases as
$u_h$ is lowered and for 
\bg\label{uhval}
u_h ~ = ~ {1.1067\over {\cal A}^{1/4}_4}
\nd
we expect the potental to have a minimum slope where the onset of {\it melting} should appear. A temperature greater 
than this is physically not possible because the string would break. Also note if we kept ${\cal A}_0 \ne 1$, 
we would have $({\cal A}_0/{\cal A}_4)^{1/4}$ in \eqref{uhval}.  

The above conclusion is certainly interesting, but may be a little 
naive because there is no strong reason to terminate the constraint \eqref{real-5} to 
${\cal A}_4$ because \eqref{newvar} would tell us that higher powers of $u_{\rm max}$ will have all the lower 
${\cal A}_n$'s as coefficients. This would make the subsequent analysis complicated, and we have to resort to 
numerical methods. 

On the other hand the above analysis does shed some light on the 
situation where the ratio ${u_{\rm max}^4\over u_h^4} > 1$. It is clear from our above calculation what happens when 
$u_h$ becomes too small: since by lowering $u_h$ there is an increase in 
${\bf u}_{\rm max}$, we are always bounded by the 
constraint: 
\bg\label{conh}
{\bf u}_{\rm max} ~\le~u_h
\nd
where the equality would lead to \eqref{uhval}. Therefore for \eqref{simpol} and \eqref{conh} to make sense, 
the string connecting quark and the antiquark should {\it break} when ${\bf u}_{\rm max}$ starts exceeding 
$u_h$ as mentioned above. 
This is the point where total melting happens, and the linear potential goes from the minimum slope $\sigma$,
where
\bg\label{minslope}
 \sigma ~\equiv ~ {V_{Q\bar Q}\over d} ~ = ~ 1 + 0.32~{\cal A}_2
\nd
to zero as soon as the temperature is increased, or alternatively $u_h$ is decreased, beyond \eqref{uhval}. In fact 
beyond certain temperature, constraint equation like \eqref{real-5} is no longer valid, and we are only left with the 
Coulombic term that eventually dies off at large distances.  
  
\subsection{Numerical analysis}

Most of the calculations that we did in the previous sections have been analytic. Under some approximations we could 
see certain important properties of quark-antiquark potentials at zero and non-zero temperatures. However as mentioned 
in the footnote earlier, truncating the series \eqref{real-5} as \eqref{quartico} may not capture the full story 
although the above toy example does give us a way to compute the melting temperature where the slope of the potential
hits a minima \eqref{minslope}. Our main conclusion of the previous section is that there exists a set of 
warp factors ${\cal A}_n$ for which the equality ${\bf u}_{\rm max} = u_h$ is valid. That means out of a large 
sets of possible backgrounds (classified by the choices of warp factors satisfying EOMs) this equality would 
select a particular subset of backgrounds that allow deconfinement and quarkonium meltings at the temperatures 
\eqref{uhval} for a subset of ${\cal A}_4$ in this approximation. 
What happens if we choose an 
arbitrary set of warp factors that do not lie in this subset?
In this section 
we will perform a numerical analysis to study the behavior of the quark-antiquark potentials for this case
at all temperatures.

For this particular choice of coefficients of the warp factor we can numerically compute the interquark distance 
$d$ in 
(\ref{D-1}), (\ref{DT-1}) and the NG action $S_{\rm
NG}^{\rm ren}$ (\ref{NG-3}),(\ref{NGT-3}), for various values of $u_{\max}$ and using this, 
plot $V_{Q\bar{Q}}$ as a function of $d$. The
analytic behavior of d and $V_{Q\bar{Q}}$ discussed in the previous sections for $u_{\rm max}$ very large and small will
turn out to be consistent with our numerical analysis although the property of melting will not be visible now. 
For simplicity we will choose the following values for the coefficients
of the warp factor: 
\bg\label{wfnumber}
{\cal A}_0~=~1,~~~~{\cal G}_0~=~1,~~~~{\cal A}_2~=~{\cal A}_4~=~{\cal G}_2~=~{\cal G}_4~=~0.24
\nd
with
$g_s\sim 0.02$ and $N_f=24$. This is indeed a reasonable choice, and hopefully satisfies EOMs despite being outside the 
required subset, because 
as we saw from F-theory, all corrections to warp factor due to
running of the $\tau$ field comes as ${\cal O}(g_s^2N_f^2)$.  Figure 7 
shows how the inter-quark distance $d$ 
varies with $u_{\rm max}$.
\begin{figure}[htb]\label{VQQ3}
		\begin{center}
\includegraphics[height=9cm,angle=-90]{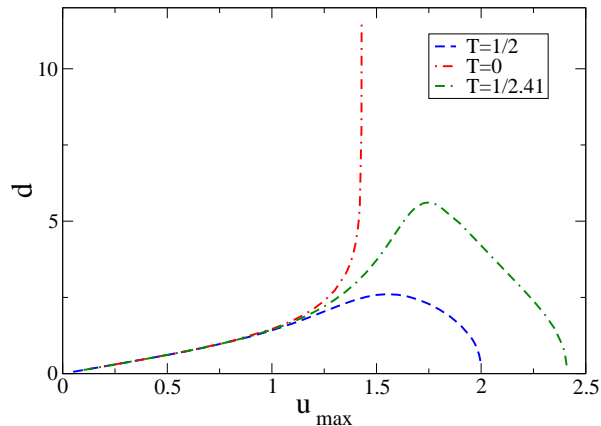}
		\caption{Inter-quark distance d as a function of $u_{\rm max}$ evaluated for 
our choice of warp factor given earlier. The red curve is the zero temperature limit. 
Here $T \equiv 1/u_h \equiv r_h$ henceforth.}
		\end{center}
		\end{figure}
For $T=0$, from the figure we see that there exists an upper bound $\bf{u}_{\rm max}$ near which $d\rightarrow \infty$. A
similar analysis also gives as $u_{\max}\rightarrow \bf{u}_{\rm max}$, $V_{Q\bar{Q}}\rightarrow \infty$.   By
increasing $u_{\rm max}$ near $\bf{u}_{\rm max}$, we can get {\it all} the values of $d$ and $V_{Q\bar{Q}}$ and using this
we can
plot $V_{Q\bar{Q}}$ as a function of $d$ as shown in  Figure 8. Note that
for large $d$, the potential grows exactly
linear with distance indicating linear confinement. This is consistent with 
our earlier analytical calculations. Additionally for small
distances, the potential behaves like a Coulomb potential as also predicted by our analysis. 		  
\begin{figure}[htb]\label{VQQ2}
		\begin{center}
\includegraphics[height=9cm,angle=-90]{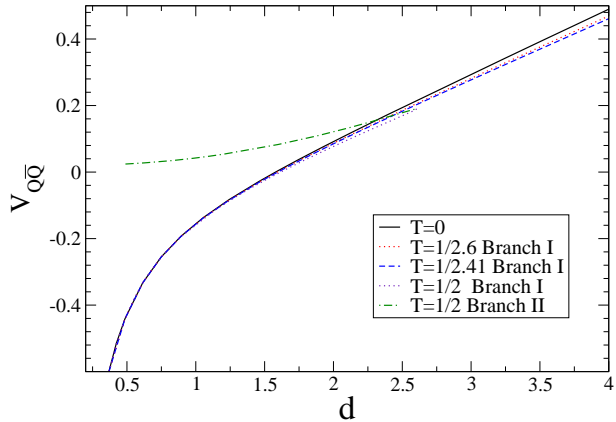}
		\caption{Quarkonium potential as a function of inter quark distance $d$ at 
various temperatures. Note the linear and the Coulombic behaviors at large and small distances respectively.}
		\end{center}
		\end{figure}
On the other hand for $T\neq 0$, again for the choice of our warp factors \eqref{wfnumber}, 
from  Figure 7 we observe that for
every $T\neq 0$ curve,
there exists a maximum value of $d$, say
$d_{\rm max}$ and therefore
for every value of $d$, there are two distinct values for $u_{\rm max}$. Such a behavior has also been observed 
in \cite{reyyee} for the AdS case and in \cite{cotrone, cotrone2} for the pure Klebanov-Strassler case.
This means for a particular
 choice of $d$ and boundary condition $u(\pm d/2)=0$, there are two $U$-shaped 
strings with two values for $u_{\rm max}$, namely $u_{\rm max,1}$ and 
 $u_{\rm max,2}$ 
with $u_{\rm max,2}> u_{\rm max,1}$. 
As $u_{\rm max,2} > u_{\rm max,1}$, the $U$-shaped string with $u_{\rm
  max,2}$ has higher energy than the one with $u_{\rm max,1}$. We have denoted 
the $U$-shaped string with $u_{\rm max,1}$ by branch
  I and the one with $u_{\rm max,2}$ by branch II in  Figure 8. It is clear from the plot that branch I 
  has lower energy than branch II. 
Thus at small $d$, the potential for branch I behaves as Coulomb potential and by comparing this to the
  zero temperature Coulomb behavior, we see that the $T\neq 0$ Coulomb potential is suppresed.   
  
  Now note that as we lower the temperatue, the value for $d_{\rm max}$ increases. Therefore 
in the limit $T\rightarrow 0$, $d_{\rm
  max}\rightarrow \infty$, which is perfectly consistent with our zero temperature curve in  Figure 7. 
  This implies as $T\rightarrow 0$, the curves in Figure 7 converge with the zero temperature curve, which in turn  
blows up at $\bf{u}_{\rm max}$. Such a behavior is not inconsistent with the high temperature case because we can view
the $T=0$ curve to 
  go straight up and never come down, resulting in a single solution for the 
$U$-shaped string for every $d$. The high temperature curves go upto some $d_{\rm max}$ and then come down. 
 For large $d < d_{\rm max}$, we have linear potential for branch 
I $-$ which is suppressed compared
  to the zero temperature curve as shown in  Figure 8. As the temperature is increased 
more is the suppression and 
  smaller is 
the value for $d_{\rm max}$. 

In Figure 8 with our choice of the warp factor coefficients the slopes of
  linear potentials and the resulting
suppressions are not very significant. For a better view of the suppressions, we present a blown up 
  version of  Figure 8 in Figure 9.
  \begin{figure}[htb]\label{VQQ1}
		\begin{center}
\includegraphics[height=9cm,angle=-90]{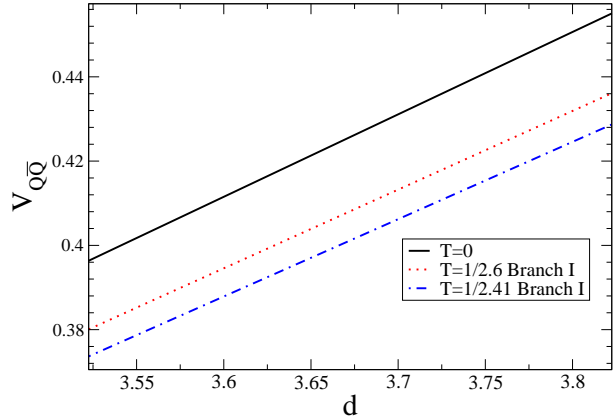}
		\caption{{Suppressions at non zero temperatures of the linear potential magnified by 
choosing a slightly different values of the coefficients of the warp factor given earlier. In this figure 
one can clearly see how with high temperatures the quark-antiquark potentials melt.}}
		\end{center}
		\end{figure}

 For $d>d_{\rm max}$, there is no real solution to the differential equation for
  $u(x)$ with boundary condition $u(\pm d/2)=0$ and thus there is no $U$-shaped string between the quarks. 
Thus for $d > d_{\rm
  max}$, the string breaks and we have deconfined quarks. Also 
as $d_{\rm max}$ decreases with increasing temperature, at high
 temperatures the quarks get screened at shorter distances $-$ which is consistent with heavy quarkonium suppression in
  thermal QCD.
  
Our numerical analysis and the plots should be instructive for a generic choice of the coefficients of the
warp factor where we again may not see the melting temperature. 
To study the generic case, first consider the zero temperature limit.
Note that the existence of
real positive $\bf{u}_{\rm max}$ is guaranteed if all $\widetilde{{\cal A}}_{n}$ in \eqref{newvar} 
are positive. However this may not be true if the original warp factor 
coefficients ${\cal A}_n$ are a {\it finite} set\footnote{As should be obvious from \eqref{newvar}, a finite 
set of ${\cal A}_n$ still implies an infinite set of $\widetilde{{\cal A}}_{n}$.}. 
This is because for a finite set
of ${\cal A}_n$'s, there will be some $\widetilde{\cal A}_n$'s which are negative and the equation 
\bg \label{mastereqn}
\frac{m+1}{2}\widetilde{\cal A}_{m+3}{\bf u}_{\rm max}^{m+3}~= ~ 1
\nd 
may not have any real positive solutions. This would imply the existence of $d_{\rm max}$ and
consequently two branches of solution. For large $d < d_{\rm max}$, we will have linear potentials with suppressions at
higher temperatures with lower values of $d_{\rm max}$. 


On the other hand, at zero temperature 
if all ${\cal A}_n > 0$, there is always a positive real $\bf{u}_{\rm max}$ and we will have linear potential at
large distances.  If some ${\cal A}_n$'s are negative,
we could have no real positive solution $\bf{u}_{\rm max}$ 
which means $0\leq u_{\rm max} \leq \infty$ as there is no black
hole horizon. In this case the behavior of d will be dominated by $d\sim u_{\rm max}$ and that of $V_{Q\bar{Q}}$ will be
dominated by $V_{Q\bar{Q}}\sim 1/u_{\rm max}$ which means $V_{Q\bar{Q}}\sim 1/d$ $-$ and we will have the 
Coulomb potential
for all $d$.

Our above numerical analysis certainly illustrates the decrease in the slope of the linear potential 
as the temperature is increased but {\it does not} show us the melting temperature. What would happen if we 
restrict our warp factor choice to the required subset of ${\cal A}_n$? In this case all the high temperature curves 
will grow linearly and will not come back, and at a certain temperature the slope of the curve will drop to 
zero. This will be the melting temperature. In this paper we will not pursue this anymore, and more details 
will be presented elsewhere.   

\section{Conclusions and Discussions}

In this paper we have tried to achieve two goals: First is to find the dual to large $N$ gauge theory that  
resembles large $N$ QCD i.e. at far IR the theory confines and at far UV
the theory shows a conformal behavior. We then extend this to high temperatures. Second, is to compute the 
heavy quarkonium potential in this theory both at zero and non-zero temperatures. We have shown that, under some
rather generic conditions, zero temperature linear confinement for heavy quarkonium states can be demonstrated. At 
high temperature, the expected deconfinement and quarkonium melting follow from our analysis. 

There are however still a few loose ends that need to be tightened to complete the full story. The first one is the 
issue of supersymmetry. Although we have shown that all the unnecessary tachyons can be removed from our picture, this 
still doesn't imply low energy supersymmetry in our model (at zero temperature). Having {\it no} supersymmetry should 
be viewed as desirable because we don't expect low energy susy in real world! However susy breaking in our model may 
trigger corrections in the potential that need to be worked out. We expect these corrections to change the coefficient 
of the linear term {\it without} generating an ${\cal O}(d^2)$ term. These corrections should be higher orders in 
$g_sN_f$, so will not change any of our conclusions presented 
here. This is because the linear potential arises from the limit where the Nambu-Goto action and the 
distance $d$ exhibit identical logarithmic divergences. This behavior should remain the same whether or not we 
have low enery susy in our model or not. Thus the linear confinement argument is particularly robust for our case. 
On the other hand the Coulombic behavior is model independent, so the coefficient of the Coulombic term should 
remain unchanged whether we take susy or non-susy models (see for example the model of \cite{zakahrov}). 
One may choose other embeddings of seven-branes, like 
\cite{kuperstein} 
or the model studied in \cite{gtpapers}, to study the quarkonium potentials at zero and non-zero temperatures.
But such choices of embeddings will not change our main conclusions.  

The second one is the issue of Higgsing that breaks the gauge group from $SU(N+M) \times SU(N+M)$ to $SU(N+M) \times 
SU(N)$ in the gauge theory side. As mentioned in sec. 2.3, the story in the dual gravity side is somewhat clearer. What
one needs is to analyse the gauge theory operators carefully that will allow the above mentioned Higgsing. We leave 
this for future work. 

The third one is to find the precise set of warp factors ${\cal A}_n$ that satisy EOMs and 
allow us to get the melting temperature for the heavy quarkonium states. In the previous section we gave a numerical
analysis with an arbitrary truncated set of warp factors that shows the decrease in the slope of the linear 
potential with increasing temperatures. Our numerical analysis certainly shows the possibility of melting, but 
doesn't tell us the melting temperature. On the other hand our analytic way of getting the melting temperature 
is not very generic. So it would be interesting to find the full set of warp factors to complete the 
story\footnote{Note that in the series of papers \cite{karsch, boschi, brambilla} similar 
screening like what we have in \eqref{liponow} is also observed using completely different techniques than ours. Our
prediction then would be that the square-root suppression that we see in \eqref{liponow} at high temperatures
is {\it universal}. Of 
course it would be interesting to figure out the Coulomb screening at high temperature also.}.  

Finally, we haven't actually computed the exact gauge fluxes on the seven and five-branes that would cancel the 
tachyons in this model. Following the works of \cite{susyrest} this may not look like a difficult task to do, at least 
for the flat background. What makes it non-trivial here is that all the branes are embedded in a {\it curved} background.
Quantisation of strings in a curved background is highly non-trivial, so it'll be rather challenging to work this 
out in full details. Nevertheless, if we restrict everything to Region 3 and away from the brane-antibrane systems, these
subtletes will not affect our results in any significant way. Happily this is the regime where most of our calculations
have been performed in this paper.

\vskip.2cm

\noindent {\bf Note}: As this draft was being written, we became aware of the work of Gaillard {\it et ~al} 
\cite{martelli} which has some overlap with this paper. See also the earlier work \cite{marmal}.

\vskip1cm

\noindent{\bf Acknowledgements}

\vskip.2cm

Its our pleasure to thank Peter Ouyang for many helpful discussions, and comments on the preliminary version of 
our draft. We would also like to thank Dongsu Bak, Niky Kamran and Omid Saremi for helpful comments. 
M. M would like to thank the 
organisers of {\it String 2010} for comments on the poster demonstration of our work. He would also like to thank
Chris Herzog, Dario Martelli, Jorge Noronha and Ashoke Sen for helpful comments. This work is supported in part by the 
Natural Sciences and Engineering Research Council of Canada, and in part by McGill University.

\appendix
\section{Complete analysis of a background configuration}

In this appendix we will, for illustration, compute the background that appears from the backreactions of the 
seven branes without any three-form fluxes but with five-form fluxes. These five-form fluxes are 
the remnant of the three-branes. So in the gauge-theory
side we will have a system of $N$ D3 branes and $N_f$ seven branes.
The excitations of these D3/D7 branes are described by a gauge theory with $SU(N)\times SU(N)$ color symmetry and 
$SU(N_f)\times SU(N_f)$ flavor symmetry. Holography dictates that the near horizon geometry sourced by these D3/D7 
branes is dual to the 
gauge theory (decoupled from gravity) which arises from the brane excitations. Of course there are two ways to 
get the gravity dual picture from the brane configuration.  
We could first obtain the geometry sourced by the 
D3/D7 system and then take the near horizon limit of it. Or first compute the near horizon 
geometry of the D3 branes, 
then place seven-branes in that geometry and finally compute the backreaction. Both this approaches are identical in 
the limit where the stack of seven-branes are separated from the stack of D3-branes. Therefore for technical simplicity,
we will adopt the latter approach. We will also put the configuration in a conifold setting to mimic what we 
did in \cite{FEP}. 

The near horizon geometry of the stack of D3 branes placed at the tip of a conifold is $AdS_5\times T^{1,1}$. 
We embed the seven-branes in this
background with the world-volume filling up four Minkowski directions plus a four-cycle in $T^{1,1}$,
and compute the backreaction. The supergravity action describing the geometry in Einstein frame is
\bg \label{action1}
S_{\rm SUGRA}=\frac{1}{2\kappa^2_{10}}\int d^{10}x \sqrt{G}\left(R+\frac{\partial_\mu \bar{\tau}\partial^\mu\tau}{2|{\rm
Im}\tau|^2}-\frac{1}{2}|\widetilde{F}_5|^2\right)+S_{\rm D7}^{\rm loc}+
\int_{\Sigma_8} C_4\wedge R_{(2)}\wedge R_{(2)}\nonumber\\
\nd
where $\tau$ is the axio-dilaton field, $\widetilde{F}_5$ is the five-form fluxes sourced by the D3 branes, 
$C_4$ is the pull-back of the four-form potential,  
$G=\sqrt{{\rm det}~G_{\mu\nu}}$ with $G_{\mu\nu}$
being the metric, $R_{(2)}$ is the curvature two-form, and $S_{\rm D7}^{\rm loc}$ is the local action for the 
seven-branes. There are additional Chern-Simons couplings of the seven-brane gauge fields to the background RR forms, 
but we will ignore them for the time being. They will appear later.


Varying the action with respect to the various fields give rise to the background equations of motion. Using our 
earlier notations, they are given by \eqref{rten} (without the ($p,q$) five-brane terms)
and \eqref{5form} (again removing the ($p,q$) five-brane contribution)
for the metric and the five-form respectively. Our metric ansatze remains
\bg\label{bhmetko}
ds^2 = {1\over \sqrt{h}}
\Big[-g(r)dt^2+dx^2+dy^2+dz^2\Big] +\sqrt{h}\Big[g(r)^{-1}g_{rr}dr^2+ g_{mn}dx^m dx^n\Big]
\nd  
as before, with $g(r)$ being the Black-Hole factor and $h$ being the warp factor that depends on all the internal 
coordinates ($r, \theta_i, \phi_i, \psi$).  To zeroth order in $g_sN_f$ we have our usual relations:
\bg\label{0gsnf}
h^{[0]} ~=~ {L^4\over r^4}, ~~~~g^{[0]} = 1-{r_h^4\over r^4}, ~~~~g^{[0]}_{rr} = 1, ~~~~
g^{[0]}_{mn}dx^m dx^n = ds^2_{T^{11}}
\nd
But in higher order in $g_sN_f$, both the warp factor and the internal metric get modified because of the back-reactions
from the seven-branes. We can write this as:
\bg\label{wfmet}
h ~ = ~ h^{[0]} ~+~ h^{[1]}, ~~~~~~ g_{rr} ~=~ g_{rr}^{[0]} ~+~ g_{rr}^{[1]}, ~~~~~~ g_{mn} ~=~ g^{[0]}_{mn} ~+~ 
g^{[1]}_{mn}
\nd
where the superscripts denote the order of $g_sN_f$. 

Using the full F-theory completion of the background as discussed in section 2.3, we know that near any one of 
the seven-branes, i.e $z\sim z_k$:
\bg
\tau~\sim ~{\rm log}(z-z_k)
\nd  
and therefore the internal metric components will typically behave as:
\bg \label{gab1}
{g}^{[1]}_{rr} ~ = ~ \sum_{i,j}~a^{[1]}_{rr,ij}~{{\rm log}^i(r)\over r^{j}}, ~~~~
{g}^{[1]}_{mn} ~ = ~ \sum_{i,j}~a^{[1]}_{mn,ij}{{\rm log}^i(r)\over r^{j-2}}
\nd 
with $a^{[1]}_{mn,ij}$ independent of $r$ but depend on the internal coordinates ($\psi,\theta_i,\phi_i$).

Now away from $z\sim z_k$, we can use similar discussion as in sec. 2.3 with axio-dilaton $\tau$ behaving exactly as
\eqref{modmap}. In that case the internal components of the metric become:
\bg \label{gab2}
{g}_{mn}^{[1]}~= ~ \sum_{i=0}^{\infty}\frac{a_{mn,i}^{[1]}}{r^{i-2}}, ~~~~~ {g}_{rr}^{[1]}~=~ 
\sum_{i=0}^{\infty}\frac{a_{rr,i}^{[1]}}{r^i}
\nd  
where again $a^{[1]}_{mn,i}$ are independent of $r$ but depend on the internal coordinates. 
 
Now to find the precise expression for $h$, we use the five-form equation \eqref{5form}. Since both the three-forms
vanish for our case, and additionally if we embed part of 
the spin connection in the seven-brane gauge connection with 
appropriate number of antiseven-branes, 
we can 
easily derive the following warp factor equation: 
\bg \label{bianchi2}
\sum_n~\frac{\partial}{\partial x^m}\left(g^{mn}g^{00}\sqrt{-{\rm det}~g_{ab}}\frac{\partial h}{\partial x^n}\right)
= ~{\rm sources}
\nd
as before;
where ($m, n$) now run over all the internal six coordinates. The generic solutions of all the above equations would be
\bg\label{hdenow}
&& h ~= ~\frac{L^4}{r^4}\left[1+\sum_{i=1}^\infty\frac{a_i(\psi,\theta_j,\phi_j)}{r^i}\right]~~~ 
{\rm  for ~~ large} ~r\nonumber\\
&& h ~= ~\frac{L^4}{r^4}\left[\sum_{j,k=0}\frac{b_{jk}(\psi,\theta_i,\phi_i){\rm log}^kr}{r^j}\right]~~~
{\rm  for ~~ small} ~r
\nd
which is of course the solutions \eqref{hde} discussed before, and away from the interpolating region, i.e Region 2.
In the next appendix we will discuss a way to determine
the coefficients $a_i$ and $b_{ij}$. Note that the above form of the large $r$ warp factor also implies that the 
effective number of colors is given by:
\bg \label{Neff}
N_{\rm eff}~= ~N~\left(1+\sum_{i=1}\frac{a_i}{r^i}\right)
\nd
Thus $N_{\rm eff}$ keeps growing with radial coordinate $r$ for all 
$a_i<0$ and $N_{\rm eff}=N$ at the bounday $r=\infty$. Alternatively, with
a change of coordinate $u = 1/r$, the metric will take the following form: 
\bg \label{reg3metnou}
ds^2&=& g_{\mu\nu} dX^\mu dX^\nu ~=~
{\cal A}_n(\psi,\theta_i,\phi_i)u^{n-2}\left[-g(u)dt^2+d\overrightarrow{x}^2\right]\nonumber\\
&+&\frac{
{\cal B}_l(\psi,\theta_i,\phi_i)u^{l}}{
{\cal A}_m(\psi,\theta_i,\phi_i)u^{m+2}g(u)}du^2+\frac{1}
{{\cal A}_n(\psi,\theta_i,\phi_i)u^{n+2}}~ds^2_{{\cal M}_5}
\nd
which is the metric \eqref{reg3met} and $ds^2_{{\cal M}_5}$ is the metric of the deformed $T^{1,1}$.
We note as before that the coefficients $C_n$ depend on the coordinates ($\psi, \theta_i, \phi_i$), the locations
of the seven-branes, 
and the number of colors and flavors, $N$ and  $N_f$ respectively. 
For all $C_n>0$, $N_{\rm eff}$ grows with decreasing $u$ with maximum value at the boundary $u = 0$. 
 


\section{Solution to Einstein equations}

One last thing that needs to be studied is the deviation of the internal metric from the usual Ricci-flat metric. We 
already hinted this in sec. 2.3, but now we will keep all the angular coordinates and see whether we can find a 
possible solution. Our aim therefore is to solve \eqref{rten}, keeping only the axio-dilaton sources \eqref{modmap}
and not the seven-branes and antiseven-branes local energy-momentum tensors. A justification for this can be easily 
provided: we are using the {\it efective} embeddings \eqref{embedding} and \eqref{ABembedding} for branes and antibranes
respectively. The axio-dilaton field \eqref{modmap} can alternatively be written as: 
\bg \label{tfield}
\tau=\sum_n {\cal A}_n z^{-n}=\sum_n {\cal A}_n r^{-n} e^{in(\phi_1+ \phi_2 -\psi)}
\left[{\rm cosec}\left(\frac{\theta_1}{2}\right){\rm
cosec}\left(\frac{\theta_2}{2}\right)\right]^n
\nd
The above behavior, as discussed earlier, is valid for all $r$ (in Region 3). Therefore 
$\tau$ has a nice Taylor series expansion and the right hand side of \eqref{rten} 
will also have a
smooth Taylor  series. As an illustration, consider  
\bg
\frac{\partial{\tau}}{\partial{r}}\frac{\partial \bar{\tau}}{\partial r} =\sum_{k,l} 
{kl {\cal A}_k{\cal A}_l\over r^{k+l+2}}
e^{i(\phi_1+\phi_2 - \psi)(k-l)}
{\rm cosec}^{k+l}\left(\frac{\theta_1}{2}\right){\rm
cosec}^{k+l}\left(\frac{\theta_2}{2}\right){\widetilde F}_{rr}(\theta_1,\theta_2)
\nd
where ${\widetilde F}_{mn}(\theta_1,\theta_2)$ depends on the partial derivative 
$\vert\partial_r{\tau}\vert^2$ . As the
source is a Taylor series in $r^{-1}$, we must have Ricci-tensor ${\widetilde R}_{rr}$ to be a Taylor series. 
This in turn would imply that the {\it unwarped} metric components should go like:
\bg
\widetilde{g}_{mn} ~= ~ \sum_{i=0}^{\infty}{a_{mni} \over r^{j}}
\nd    
where $j = i-2$ or $j = i$ depending on whether we are choosing the radial direction or not (see sec. 2.3 for details).
The coefficients $a_{mni} \equiv a_{mni}(\theta, \phi, \psi)$ i.e functions of all the internal coordinates. 

Now to explicitely solve for the partial differential equations involving  $a_{mni}$
we observe that the right hand side of \eqref{rten} as a function of internal coordinates
$(\psi, \phi_i, \theta_i)$ is of the form  $e^{i(\phi_1+\phi_2-\psi)}$ and 
${\rm cosec}\left(\frac{\theta_i}{2}\right)$. 
So the most general form for  $a_{mni}$
should also involve these variables\footnote{We thank Niky Kamran for helpful comments.}.
Thus we introduce variables:
\bg
\widetilde{z}_1~= ~{\rm cos}\left(\frac{\theta_1}{2}\right)+i{\rm sin}\left(\frac{\theta_1}{2}\right)\nonumber\\
\widetilde{z}_2~=~{\rm cos}\left(\frac{\theta_2}{2}\right)+i{\rm sin}\left(\frac{\theta_2}{2}\right)
\nd 
and using which we 
make the following ansatz for $a_{mni}$:
\bg
a_{mn\alpha} = a_{mn0}+
\sum_{k,l,k+l=\alpha}e^{i(\phi_1 + \phi_2 - \psi)(k-l)}\Bigg[{\widetilde{a}^{ij}_{mn}\over 
\widetilde{z}_i\widetilde{z}_j^*} 
+{\widetilde{a}^{ijop}_{mn}\over \widetilde{z}_i\widetilde{z}_j^*\widetilde{z}_o\widetilde{z}_p^*}
....+~{\widetilde{a}^{i_1i_2..i_{2n}}_{mn}\over 
\widetilde{z}_{i_1}\widetilde{z}_{i_2}^*....\widetilde{z}_{i_{2\alpha-1}}\tilde{z}_{i_{2\alpha}}^*}\Bigg]\nonumber\\
\nd
where summation over repeated index is assumed. 
Note that
with this ansatz written in terms of coordinate $\widetilde{z}_i$, the differential equations 
in \eqref{rten} become a set
of algebraic equations because ${\rm sec}^n(\theta_i),{\rm cosec}^n(\theta_i)$ etc. 
can be written as a linear combinations of
$\widetilde{z}_i$. 
Then it is straight forward to equate coefficients of various powers of $\widetilde{z}_i$ and obtain the {\it constant}
coefficeints $\widetilde{a}_{i_1i_2..i_{2n}}$. 
This gives us the precise procedure to obtain the exact solutions for all $a_{mni}$.  
In Region 2 we have to be more careful because there are, in addition to the axio-dilaton, other sources. But since 
all our calculations are restricted to Region 3, we need not worry too much about the non-Ricci flatness of the 
internal manifold.


\end{document}